\documentclass[acmtog,review=false]{acmart}
\def\CiteSelf/{{yes}}
\def\ProvideEmail/{{yes}}

\copyrightyear{2020}
\acmYear{2020}
\setcopyright{acmcopyright}
\acmJournal{TOG}
\acmYear{2020}\acmVolume{39}\acmNumber{4}\acmArticle{145}\acmMonth{7}
\acmDOI{10.1145/3386569.3392458}

\usepackage{nth}
\usepackage{siunitx}  

\DeclareMathOperator{\sgn}{sgn}  
\DeclareMathOperator{\denom}{denom}  

\usepackage{mathtools}
\DeclarePairedDelimiter\ceil{\lceil}{\rceil}

\def\NVpr/{{\tt NV\_path\_\-rendering}}
\def\NVprs/{{\tt NV\_path\_\-rendering}'s}
\def\NVbea/{{\tt NV\_blend\_equation\_\-advanced}}
\def\KHRbea/{{\tt KHR\_blend\_\-equation\_\-advanced}}
\def\NVbb/{{\tt NV\_blend\_barrier}}
\def\NVfms/{{\tt NV\_framebuffer\_mixed\_samples }}
\def\ARBtms/{{\tt ARB\_texture\_multisample}}
\def\Bezier{B\'ezier }		

\ifdefined\CiteSelf
\newcommand{\mycite}[1] {\cite{my-#1}}
\else
\newcommand{\mycite}[1] {\cite{anon-#1}}
\fi

\newcommand{\urlwofont}[1]
{
\urlstyle{same}\url{#1}
}

\begin{document}

\title[Polar Stroking]{Polar Stroking: New Theory and Methods for Stroking Paths}

\author{Mark J. Kilgard}
\affiliation{%
\institution{NVIDIA}
\city{Austin}
\country{USA}
}
\ifdefined\ProvideEmail
\email{mjk@nvidia.com}
\fi

\renewcommand{\shortauthors}{Mark Kilgard}

\begin{abstract}

Stroking and filling are the two basic rendering operations on paths in vector graphics.
The theory of filling a path is well-understood in terms of contour integrals and winding numbers, but
when path rendering standards specify stroking, they resort to the analogy of painting pixels with a brush that traces
the outline of the path.  This means important standards such as PDF, SVG, and PostScript lack
a rigorous way to say what samples are inside or outside a stroked path.  Our work fills this gap with a principled theory of stroking.

Guided by our theory, we develop
a novel {\em polar stroking} method to render stroked paths robustly with an intuitive way to bound the
tessellation error without needing recursion.
Because polar stroking guarantees small uniform steps in tangent angle, it provides an
efficient way to accumulate arc length along a path for texturing or dashing.  While this paper focuses on developing the theory
of our polar stroking method, we have successfully implemented our methods on modern programmable GPUs.

\end{abstract}




\begin{teaserfigure}
\centering
\includegraphics[width=\columnwidth]{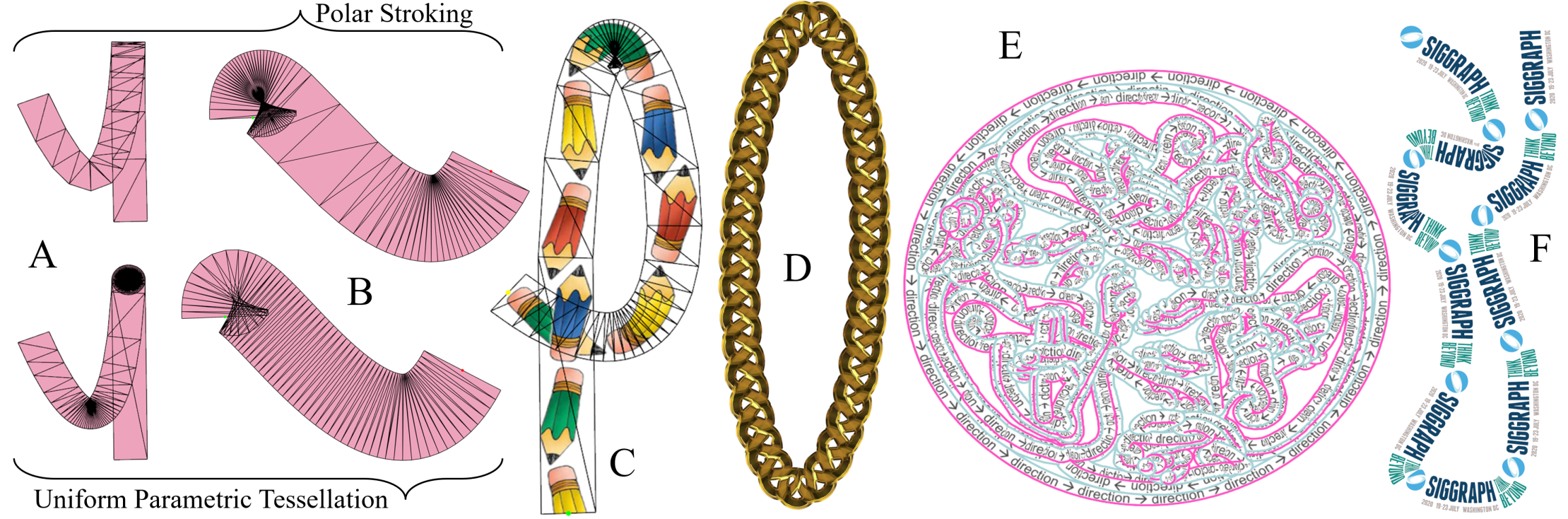}
\caption{\normalfont Polar stroking samples: {\bf A} cubic \Bezier segment with a cusp rendered properly with polar stroking while uniform parametric tessellation has no cusp, both using 134 triangles; {\bf B} polar stroking improves the facet angle distribution compared to uniform tessellation, both using 126 triangles; {\bf C} arc length texturing; {\bf D} ellipse drawn as just 2 conic segments, one external; {\bf E} complex cubic \Bezier path (5,031 path commands, 29,058 scalar path coordinates) with cumulative arc length texturing; {\bf F} centripetal Catmull-Rom spline. \label{fig:teaser}}
\end{teaserfigure}

\maketitle


\section{Introduction}

Vector graphics standards such as
PDF \cite{PDF-Spec},
SVG \cite{SVG-Spec},
PostScript \cite{PostScript-Spec},
PCL \cite{PCL-Spec},
HTML5 Canvas \cite{Canvas-Spec},
and XPS \cite{XPS-Spec} support two basic rendering operations on paths: stroking and filling.

The intuition of stroking a path is like a child drawing in a coloring book by ``tracing over the lines'' and treating each path as the outline to trace.
Filling a path is like ``coloring inside the lines.''

The stroking operation on paths---mandated and specified by all the listed standards above---lacks a mathematically grounded theory to define what stroking
means.  To remedy this situation, we aim to provide a principled theory for stroking and show our theory motivates robust, useful, and GPU-amenable methods for stroking.

\subsection{A Quick Theory of Path Filling}
\label{sec:quick_theory}

We first review the theory of path filling to show filling indeed has a principled theory---in contrast to path stroking.

When a path is filled, pixels ``inside'' the path get shaded and composited.  At first glance, path filling sounds simple, but a path can be arbitrarily complex.  It can be empty, concave (perhaps extremely so), intersect itself, contain multiple closed regions (some of which wind clockwise while others the reverse), contain curved sections as well as straight ones, and may be degenerate in various ways (exhibiting cusps or closed regions with no interior).  So a computer's decision whether a pixel is ``inside or not'' may seem quite involved---even ill-defined, but good theory turns path filling into an unambiguous, well-defined, and ultimately rote rendering operation. 

Vector graphics turns path filling into a rigorously defined operation by adopting the theory of contour integrals from complex analysis.
Appendix~\ref{sec:fill-theory} reviews this theory and its winding number concept.

Graphics practitioners have long appreciated and richly mined discrete versions of this theory as a sound
basis for efficient filled path rasterization algorithms.
The details are outside our scope, so we cite just a few examples.
Lane et al.~\shortcite{Lane:1983:AFR:357323.357326} rasterized concave polygons this way.
Corthout and Pol~\shortcite{pharos-thesis} formalized use of discrete curved contours for rasterization and
Fabris et al.~\shortcite{fabris1997efficient} improved its efficiency.
Kilgard and Bolz~\shortcite{KilgardBolz2012} combined stencil buffer methods with insights from
Loop and Blinn~\shortcite{Loop:2005:RIC:1186822.1073303} to fill paths
efficiently using GPU rasterization and shading.  Scanline rasterizers for filling paths \cite{Ackland:1981:EFA:1963620.1963624,Kallio} practice this theory in 1D on rows of pixels.

\subsection{Good Theory Would Benefit Stroking Too}
\label{sec:good-theory}

While path filling is explicitly specified to depend on a pixel's winding number, no such rigorous underpinning
exists for path stroking in any of the specifications of major vector graphics standards.

For example, the PDF standard~\shortcite{PDF-Spec} states its stroke operator ``shall paint a line along the current path'' and ``shall follow each straight or curved segment in the path, centered on the segment with sides parallel to it.''  The PDF standard's description is intuitive in its appeal to a painting metaphor.  However, a metaphor is insufficient to reason about what pixels should and should not be covered by any particular stroked path segment.

We use elements from the differential geometric theory of curves to mathematically formulate the problem of stroking a path segment. We define stroking using the concept of offset curves and take care to handle points where the derivative goes to zero (cusps) by explicit provisions/alternative path definitions. The formalization allows us to define a predicate for the stroked region and develop robust, useful, GPU-amenable methods for stroking.

\subsection{Contributions and Organization}

Our contributions are:
\begin{itemize}

\item A theory of path stroking we call {\em polar stroking} that, for the first time, provides a mathematically grounded formulation of
the path stroking operation consistent with the best consensus implementations of major vector graphics standards.

\item A nonrecursive and GPU-amenable method, based on our theory, to tessellate a stroked path by making small uniform steps in tangent angle and thereby tightly and intuitively approximating the path's stroked region.  Joins, caps, and path segments are all tessellated in a single, unified way.

\item A method for efficient arc length computation along stroked paths to harness for dashing and arc length texture mapping of stroked paths.

\end{itemize}

After this introduction, we review prior work in Section~\ref{sec:prior-work}.
Section~\ref{sec:theory} explains our new theory of path stroking.
Starting from our theory, Section~\ref{sec:method} develops our polar stroking method.
Section~\ref{sec:arc_length} explains how our polar stroking facilitates practical cumulative arc length computations along a path for arc length texturing and dashing. 
Section~\ref{sec:experimental-results} compares polar stroking to uniform parametric tessellation and existing real-world software implementations.
Section~\ref{sec:limitations} reviews limitations of our methods.
Section~\ref{sec:conclusion} concludes.  Figure~\ref{fig:teaser} demonstrates various polar stroking results.


\section{Prior Work}
\label{sec:prior-work}

\subsection{Not Classic Curve and Line Rendering}

We distinguish path stroking from the classic rasterization algorithms for line \cite{bresenham1965algorithm}
and curve \cite{pitteway1967algorithm} rendering that we term ``connect the pixels'' approaches.
In these algorithms each line or curve segment is rendered as its own distinct primitive.  The idealized line or curve is 1D, even
when such segments are rendered wide or antialiased.  What width these lines have is expressed in pixel units.

In contrast, a stroked path defines a {\em 2D region} orthogonally offset from the path's generator curve by half the path's stroke width.  This width is specified in the same coordinate space
as the path's control points.  Sequences of path segments are connected by joins and start and stop with caps.  Paths
can be arbitrarily complicated in the ways listed in Section~\ref{sec:quick_theory}, and all those complications (cusps, etc.) must be handled properly.
Pixel-space line primitives can be stippled, but the dashing of paths is considerably more complicated, taking
place in the path's own coordinate system and operating on curved paths.

\subsection{Path Rendering's Stroking Operation}

The foundational work of Warnock and Wyatt~\shortcite{Warnock:1982:DIG:800064.801297} outlines
a complete device-independent vector graphics system.  Their paper describes an operation whereby a brush follows a trajectory
to generate a shape that can then be drawn using filling.  The paper never uses the terms {\em stroke} or {\em path}
but by converting trajectories to shapes to be filled, their system foresees the path filling and stroking
operations essential to path rendering.

In this same time frame, Turner~\shortcite{Whitted:1983:ALD:800059.801144} and Hobby~\shortcite{Hobby85} explained {\em brush extrusion} approaches whereby a logical brush or pen tip of some shape is dragged along some trajectory and whatever pixels are ``swept out'' by the brush or pen tip are considered part of the rasterized region.  The brush shape and size is specified in pixel-space units.

PostScript arrived in 1984 \cite{PostScript-Spec} providing both {\em stroke} and {\em fill} operators 
on paths with support for stroke width, dashing, joins, and caps.  PostScript-style stroking is our focus.

\subsection{Stroking as a Brushing Operation}
\label{sec:brush-trajectory}

Corthout and Pol~\shortcite{pharos1991} were the first to formulate a rigorous stroking definition based on the Minkowski sum of a trajectory and a brush and
used it to reason about algorithms for stroking PostScript.
Fabris et al.~\shortcite{FabrisSilvaForrest98} further refined the underlying theory to implement a more efficient algorithm.

However this model does not capture the path stroking behavior of PostScript and similar standards.
Recall the PDF standard's phrasing ``paint a line ... {\em centered} on the segment {\em with sides parallel to the segment}'' (emphasis added).  This phrasing implies stroking must somehow depend on the gradient of each path segment.  However the brush-trajectory formulation ignores the gradient.

The brush-trajectory model ``stamps'' the brush pattern all along the trajectory and its neighborhood whereas path rendering standards have a wide-but-thin ``pen tip'' that sweeps the trajectory orthogonal to the trajectory's gradient.  While a circular brush generates identical coverage to a path segment with round caps, path rendering standards support cap and joins styles other than round, in which case the coverage will {\em not} match PDF or other standards, particularly at caps, joins, and the start and end of a segment.

As a practical matter, the brush-trajectory model's coverage is difficult to transform into a tessellation of triangles or other geometric primitives suited for efficient GPU rasterization.
While the brush-trajectory model has a rigorous formulation, we assess it does not meet our goal of matching the stroking behavior expected by existing path rendering standards.

\subsection{Path Stroking in Practice}

We survey established approaches to implement path stroking.

\subsubsection{Render a Filled Region Approximating the Stroked Region}

The description of the stroke operator by Gosling et al.~\shortcite{NeWS-Book} indicates that early on,
stroking was implemented by generating a fillable region corresponding to the stroked region of a path
and then drawing that derived fillable region.  Other recent path rendering systems explicitly state
they take this approach \cite{
Ganacim:2014:MVG:2661229.2661274,
Li:2016:EGP:2980179.2982434,
doi:10.1111/cgf.13622}.

An approximation strategy such as
Tiller and Hanson~\shortcite{Tiller:1984:OTP:1299971.1300480}
is necessary for curved segments.
This approximation is made difficult because the polynomial order of the stroked boundary of a path segment with a \nth{2} or \nth{3} order curve is substantially higher, \nth{6} or \nth{10} order respectively in general \cite{Farouki:1990:APP:87526.87544}.
This approach is subject to defects where stroked segments
overlap in ways so that the net result is a zero winding number for a sample that should technically be in the stroke,
thereby dropping coverage that should properly be part of the stroked region.
If, in order to avoid this, individual segments are rendered in isolation and antialiased, conflation artifacts are likely.

\subsubsection{Recursive Conversion of Stroked Paths to Polygons}

This approach recursively splits curved segments into smaller segments until sufficiently straight and then
converts the resulting sequence of nearly straight segments into a quadrilateral strip.  Care must be taken
at cusps and near-cusps of cubic \Bezier segments and other degenerate segments.  The Skia \cite{Skia-WebSite} and Anti-Grain Geometry \cite{Anti-Grain-WebSite}
renderers do this.  As this approach is recursive, it maps poorly to GPU tessellation.

\subsubsection{Approximation to Stroked Quadratic \Bezier Hulls}

Ruf~\shortcite{Ruf:2011:IBR:2018323.2018346} shows a means to construct a conservative bounding region around the offset regions of quadratic \Bezier curves so that point containment
queries with respect to a stroked quadratic \Bezier segment can be limited to inside the bounding region.
Kilgard and Bolz~\shortcite{KilgardBolz2012}
take this further by handling cubic \Bezier segments and arcs by approximating them with quadratic \Bezier splines
and moving the point containment queries into a fragment shader for GPU acceleration.

\subsection{Not Non-Photorealistic Rendering Stroking}

Techniques for Non-Photorealistic Rendering (NPR) use stroke, brush, and pen metaphors to create artistic
effects; surveys by
Hertzmann~\shortcite{Hertzmann:2003:TSS:858619.858653} in particular and
Kyprianidis et al.~\shortcite{Kyprianidis:2013:SAT:2478571.2479163} more recently explore various stroke-based techniques for NPR\@.  While stroke-based NPR techniques and vector graphics both share the term ``stroke'' and have similar
conceptual underpinnings, we address the specific stroking operation on paths found in
vector graphics standards rather than what NPR techniques broadly call stroking.


\section{Theory of Path Stroking}
\label{sec:theory}

\subsection{Path Preliminaries}
\label{sec:path-math-prelim}

\begin{figure}
\centering
\includegraphics[width=\columnwidth]{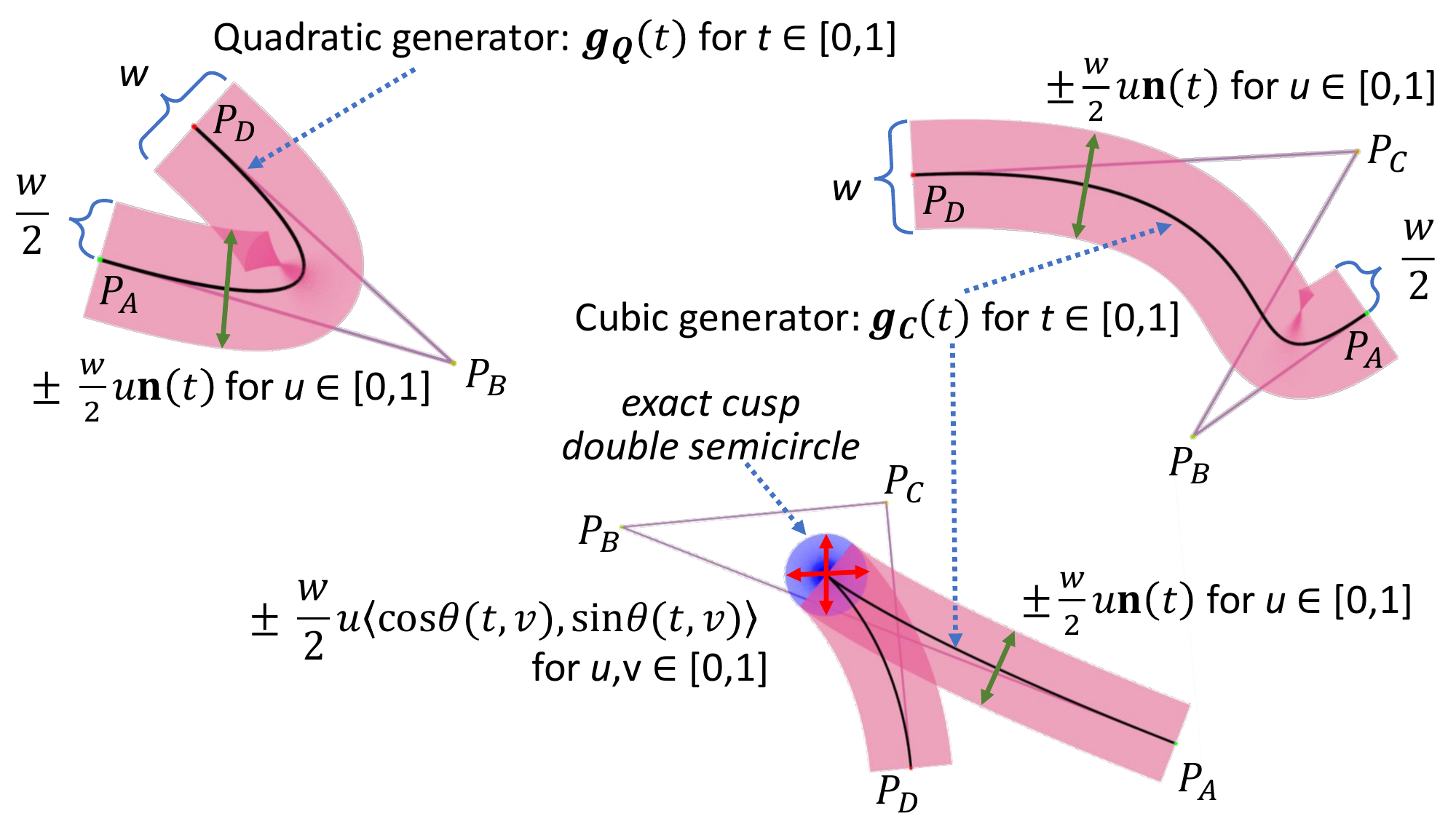}
\caption{\normalfont Geometric interpretation of Equations~\ref{eq:NaiveStroking} and \ref{eq:ConventionalStroking} for
stroking path segments shown applied to three example \Bezier segments: quadratic  ({\em upper left}), serpentine cubic  ({\em upper right}),
and exact cusp cubic ({\em bottom}).
The pink region is the stroked region according to 
Equation~\ref{eq:NaiveStroking} that fails to include the cusp's double semicircle.
The black curve within each pink stroked region is each segment's generator curve.
The pink+blue regions correspond to the stroked region according to Equations~\ref{eq:ConventionalStroking} and \ref{eq:polar-stroking}.
Green double arrows show the stroke widening term of Equation~\ref{eq:NaiveStroking}.  The red double arrows show
the cusp semicircle term of Equation~\ref{eq:ConventionalStroking}.
\label{fig:stroked-region}}
\end{figure}

A path is a sequence of $n$ path segments.  Each segment $i=1...n$ is defined by the locus of $(x,y)$ positions generated by a parametric generator curve $\mathbf{g}_{f,i}(t)$ assuming $f \in \{ C, Q, K, L \}, t \in [0,1]$.

The form $f$ of a segment selects among four parametric equations: 
cubic \Bezier ($C$), quadratic \Bezier ($Q$), conic ($K$), and linear ($L$), with each form defined in Table~\ref{tab:path-segments}
where $\mathbf{P}_A$, $\mathbf{P}_B$, $\mathbf{P}_C$, and $\mathbf{P}_D$ are $(x,y)$ control points and $w_{B}$ is a homogeneous coordinate associated with control point $\mathbf{P}_B$.  Each segment in a path has its own associated control point coordinates.  In practice segments are typically connected into splines.

\begin{table*}[t]
\caption{\normalfont Table of path segment forms in vector graphics standards.}
\begin{tabular}{ | p{3.8em} || p{18.28em} | p{24.42em} | p{3.1em} |}
\hline
{\bf Segment \newline form} &
{\bf Generator \newline curve function,} $g_f(t)$  &
{\bf Initial normalized gradient,} $\widehat{\nabla} \, \mathbf{g}(0)$ \newline {\bf Terminal normalized gradient,}  $\widehat{\nabla} \, \mathbf{g}(1)$ &
{\bf \hspace{1em} \newline Figures} \\
\hline
Cubic \newline \Bezier & 
$\begin{aligned}[t] 
\mathbf{g}_{C}(t) &={} (1-t)^3 \, \mathbf{P}_A + 3 (1-t)^2 t \, \mathbf{P}_B + \\
                  & {} \hspace{1.5em} 3 (1-t) t^2 \, \mathbf{P}_C + t^3 \, \mathbf{P}_D
\end{aligned}$
 & 
$\begin{aligned}[t]
\widehat{\nabla} \, \mathbf{g}_C(0) =
\begin{cases}
\widehat{\mathbf{P}_B - \mathbf{P}_A}, & \text{ if } ||\mathbf{P}_B - \mathbf{P}_A||>0 \\
\widehat{\mathbf{P}_C - \mathbf{P}_A}, & \text{ else if } ||\mathbf{P}_C - \mathbf{P}_A||>0 \\
\widehat{\mathbf{P}_D - \mathbf{P}_A}, & \text{ otherwise } 
\end{cases}  \\
\widehat{\nabla} \, \mathbf{g}_C(1) =
\begin{cases}
\widehat{\mathbf{P}_D - \mathbf{P}_C}, & \text{ if } ||\mathbf{P}_D - \mathbf{P}_C||>0 \\
\widehat{\mathbf{P}_D - \mathbf{P}_B}, & \text{ else if } ||\mathbf{P}_D - \mathbf{P}_B||>0 \\
\widehat{\mathbf{P}_D - \mathbf{P}_A}, & \text{ otherwise } 
\end{cases}
\end{aligned}$
  & \ref{fig:loop}, \ref{fig:serpentine}, \newline \ref{fig:serpentine_2_inflections}, \ref{fig:cusp}   \\ 
\hline
Quadratic \newline \Bezier
& 
$\begin{aligned}[t] 
\mathbf{g}_{Q}(t) &= (1-t)^2 \, \mathbf{P}_A + 2 (1-t) t \, \mathbf{P}_B + t^2 \, \mathbf{P}_D  
\end{aligned}$
& 
$\begin{aligned}[t]
\widehat{\nabla} \, \mathbf{g}_Q(0) &=
\begin{cases}
\widehat{\mathbf{P}_B - \mathbf{P}_A}, & \text{ if } ||\mathbf{P}_B - \mathbf{P}_A||>0 \\
\widehat{\mathbf{P}_C - \mathbf{P}_A}, & \text{ otherwise } 
\end{cases}
\\
\widehat{\nabla} \, \mathbf{g}_Q(1) &=
\begin{cases}
\widehat{\mathbf{P}_C - \mathbf{P}_B}, & \text{ if } ||\mathbf{P}_C - \mathbf{P}_B||>0 \\
\widehat{\mathbf{P}_C - \mathbf{P}_A}, & \text{ otherwise } 
\end{cases}
\end{aligned}$
& \ref{fig:quadratic}, \ref{fig:external_parabola} \\ 
\hline
Conic
& 
$\begin{aligned}[t] 
\mathbf{g}_{K}(t) &= \frac{(1-t)^2 \, \mathbf{P}_A + 2 (1-t) t \, w_{B}  \, \mathbf{P}_B + t^2 \, \mathbf{P}_D}
                      {(1-t)^2 \, + 2 (1-t) t \, w_{B}  + t^2 }
\end{aligned}$
& 
$\begin{aligned}[t]
\widehat{\nabla} \, \mathbf{g}_K(0) &=
\begin{cases}
\sgn(w_B) \, \widehat{\mathbf{P}_B - \mathbf{P}_A}, & \text{ if } ||\mathbf{P}_B - \mathbf{P}_A||>0 \land w_B \neq 0 \\
\hphantom{\sgn(w_B)} \, \widehat{\mathbf{P}_C - \mathbf{P}_A}, & \text{ otherwise } 
\end{cases}
\\
\widehat{\nabla} \, \mathbf{g}_K(1) &=
\begin{cases}
\sgn(w_B) \, \widehat{\mathbf{P}_C - \mathbf{P}_B}, & \text{ if } ||\mathbf{P}_C - \mathbf{P}_B||>0 \land w_B \neq 0 \\
\hphantom{\sgn(w_B)} \, \widehat{\mathbf{P}_C - \mathbf{P}_A}, & \text{ otherwise } 
\end{cases}
\end{aligned}$
&
\ref{fig:internal_ellipse}, \ref{fig:external_ellipse} \newline 
\ref{fig:internal_hyperbola}, \ref{fig:external_hyperbola} 
  \\
\hline
Line & 
$\begin{aligned}[t] 
\mathbf{g}_{L}(t) &= (1-t) \,\mathbf{P}_A + t \, \mathbf{P}_D
\end{aligned}$
& 
$\begin{aligned}[t]
\widehat{\nabla} \, \mathbf{g}_L(0) &= \widehat{\mathbf{P}_B - \mathbf{P}_A}
\\
\widehat{\nabla} \, \mathbf{g}_L(1) &= \widehat{\mathbf{P}_B - \mathbf{P}_A}
\end{aligned}$
& \ref{fig:line_segment} \\ 
\hline
\end{tabular}
\label{tab:path-segments}
\end{table*}

The conic equation $\mathbf{g}_{K}$ uses the so-called {\em normal parameterization} \cite{Piegl:1995:NB:208469} of a rational quadratic \Bezier segment, known to be sufficient to represent any conic segment \cite{lee1987rational}; we place no restrictions on $w_{B}$, allowing $w_B$ to be both zero and negative (further explained in Section~\ref{sec:angle-ranges}) and allowing for external elliptical, hyperbolic, and parabolic segments \cite{DrawingCircles}.

Path rendering standards use arc segments rather than conic segments.  For example, SVG parameterizes elliptical arcs using an {\em endpoint parameterization} \cite[\href{https://www.w3.org/TR/SVG11/implnote.html\#ArcImplementationNotes}{{\em Elliptical arc implementation notes}}]{SVG-Spec}.  All such arcs can be transformed into an equivalent $\mathbf{g}_{K}$ form.  While arc segments are more intuitive for artists creating path content, conic segments are compact, more general, more efficient to evaluate, and easier to reason about.

These four forms of path segments are the only ones needed by path rendering standards so we restrict our focus to them.  All four are smooth functions.  Linear transformation of their homogeneous control points is equivalent to the same transformation applied to points belonging to each segment's locus.

Figures~\ref{fig:loop}, \ref{fig:serpentine}, \ref{fig:serpentine_2_inflections}, and \ref{fig:cusp} illustrate $\mathbf{g}_{C}$ with topologically varied configurations of stroked cubic \Bezier segments.
Figure~\ref{fig:quadratic} illustrates $\mathbf{g}_{Q}$ forming a stroked quadratic \Bezier segment.  Figures~\ref{fig:external_parabola}, \ref{fig:internal_ellipse}, \ref{fig:external_ellipse}, \ref{fig:internal_hyperbola}, and \ref{fig:external_hyperbola} illustrate $\mathbf{g}_{K}$ forming various stroked conic  segments.  Figure~\ref{fig:line_segment} illustrates $\mathbf{g}_{L}$ forming a stroked line segment.  The specific tessellation shown for each stroked segment in each of these figures is generated with the method of Section~\ref{sec:method}.

The gradient of $\mathbf{g}(t)$ with respect to $t$ is denoted $\mathbf{g'}(t)$ or simply $\mathbf{g'}$.  The unit-length tangent $\mathbf{t}$, unit-length normal $\mathbf{n}$, and signed curvature $\kappa$ at $t$ are defined as
\[
\mathbf{t} = \frac{\mathbf{g'}}{||\mathbf{g'}||},
\hspace{2em}
\mathbf{n} = \mathbf{t} \times \mathbf{z},
\hspace{2em}
\kappa = \frac{(\mathbf{g'} \times \mathbf{g''}) \cdot \mathbf{g}}{||\mathbf{g'}||^3}
\]
where we define $\mathbf{z}=\mathbf{n} \times \mathbf{t}$ to form a unit vector perpendicular to the plane of the curve, assuming $||\mathbf{g'}||$ is nonzero.

The graph of a gradient such as $\mathbf{g'}(t)$ is known as a {\em hodograph}.  To the right of each stroked segment in Figures~\ref{fig:loop} through \ref{fig:line_segment} is the segment's hodograph on a polar plot. 

\subsection{Formulating Path Stroking}

\subsubsection{Stroking Expressed with Offset Curves}

Offset curves \cite{Farouki:1990:APP:87526.87543} depend on the gradient of their generator curve and so are better suited  than the brush-trajectory model (Section~\ref{sec:brush-trajectory}) to formulate path stroking consistent with path rendering standards.

Given a plane curve $\mathbf{g}(t)$ with a {\em regular} parameterization on $t \in [0,1]$---known as the generator curve---the offset curve to $\mathbf{g}(t)$ at a distance $d$ is defined by
\[
\mathbf{g}_o(t) = \mathbf{g}(t) + d \, \mathbf{n}(t) \; \text{ for } t \in [0,1]
\]
where $\mathbf{n}(t)$ is the unit normal to $\mathbf{g}(t)$ at each point.

We can define the stroked region of a path segment with a generator curve $\mathbf{g}(t)$ as the locus of points defined by
\begin{equation}
\mathbf{s}_{w}(t,u) = \mathbf{g}(t) \pm \frac{w}{2} \, u \, \mathbf{n}(t) \; \text{ for } t, u \in [0,1]
\label{eq:NaiveStroking}
\end{equation}
where $w$ is the stroke width.
Figure~\ref{fig:stroked-region} provides a geometric interpretation of Equation~\ref{eq:NaiveStroking}.
Observe $\mathbf{s}_{w}$ faithfully captures the PDF specification's phrasing (quoted in Section~\ref{sec:good-theory}) that a stroked region is {\em centered} because of $\pm \frac{w}{2}$ and {\em parallel to} the generator curve because the stroke boundary is offset by the normal $\mathbf{n}(t)$.

Because $\mathbf{s}_{w}$ allows simultaneous negative and positive offset distances, we may relax the previously stated regularity restriction on $\mathbf{g}(t)$ when considering its stroked region $\mathbf{s}_{w}$ because any discontinuities introduced by an abrupt reversal of the normal vector when $||\mathbf{g'}(t)||=0$ do not affect the stroked region's continuity.

\begin{figure}
\centering
\includegraphics[width=\columnwidth]{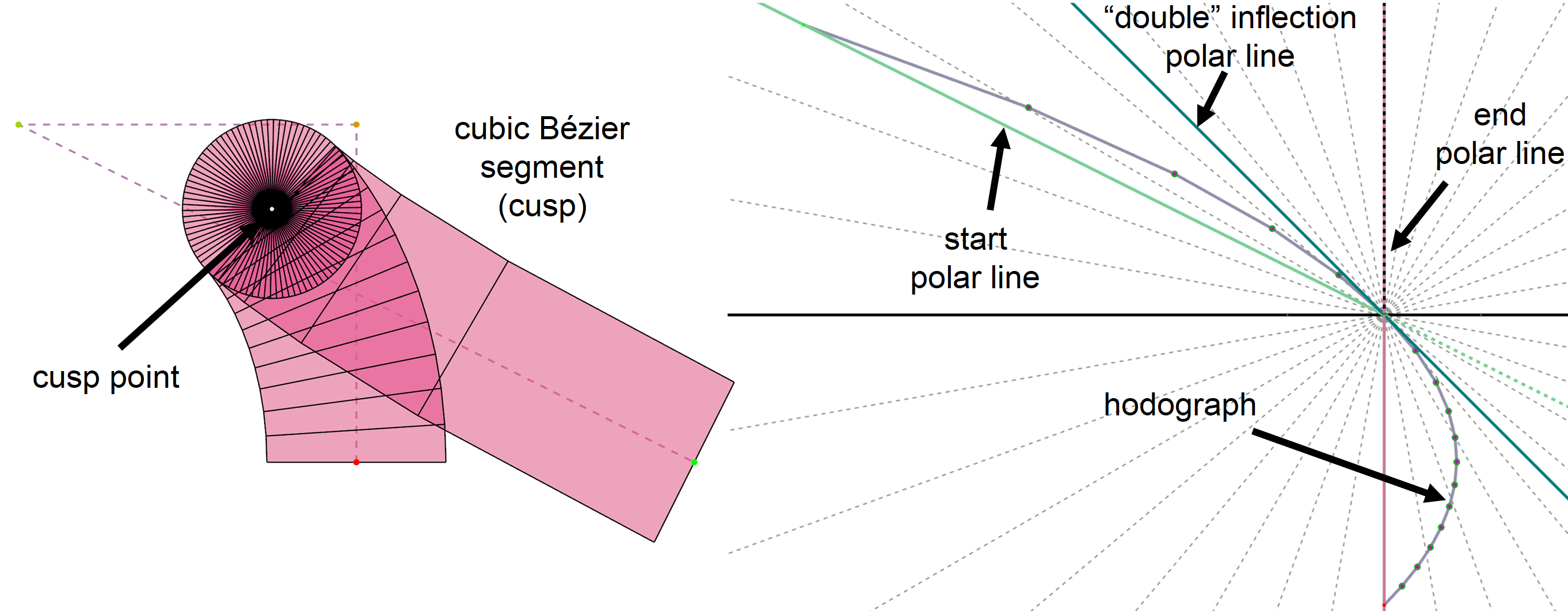}
\caption{\normalfont Cubic \Bezier segment in cusp configuration with its hodograph. \label{fig:cusp}}
\end{figure}

This relaxation is important as path rendering standards place no restrictions on path segments to guarantee regularity.  It is straightforward to specify a nondegenerate cubic \Bezier segment with an exact cusp.  See Figure~\ref{fig:cusp} for an example.  Various degenerate path segments may also induce cusps.  

To reason about the rasterized coverage of $(x,y)$ pixels with respect to a stroked path, we express $\mathbf{s}_{w}$ as a support predicate $\mathbf{s}_{w}(x,y)$ defined as
\[
\mathbf{s}_{w}(x,y) =
\begin{cases}
1, \; \text{ if } \exists_{(t,u) \in [0,1]} : (x,y) = \mathbf{s}_{w}(t,u) \\
0, \; \text{ otherwise}
\end{cases}
\]

While much closer than the brush-trajectory model to the behavior expected by path rendering standards, $\mathbf{s}_{w}$ still does not fully conform with established stroking expectations as we now explore.


\begin{figure}
\centering
\includegraphics[width=\columnwidth]{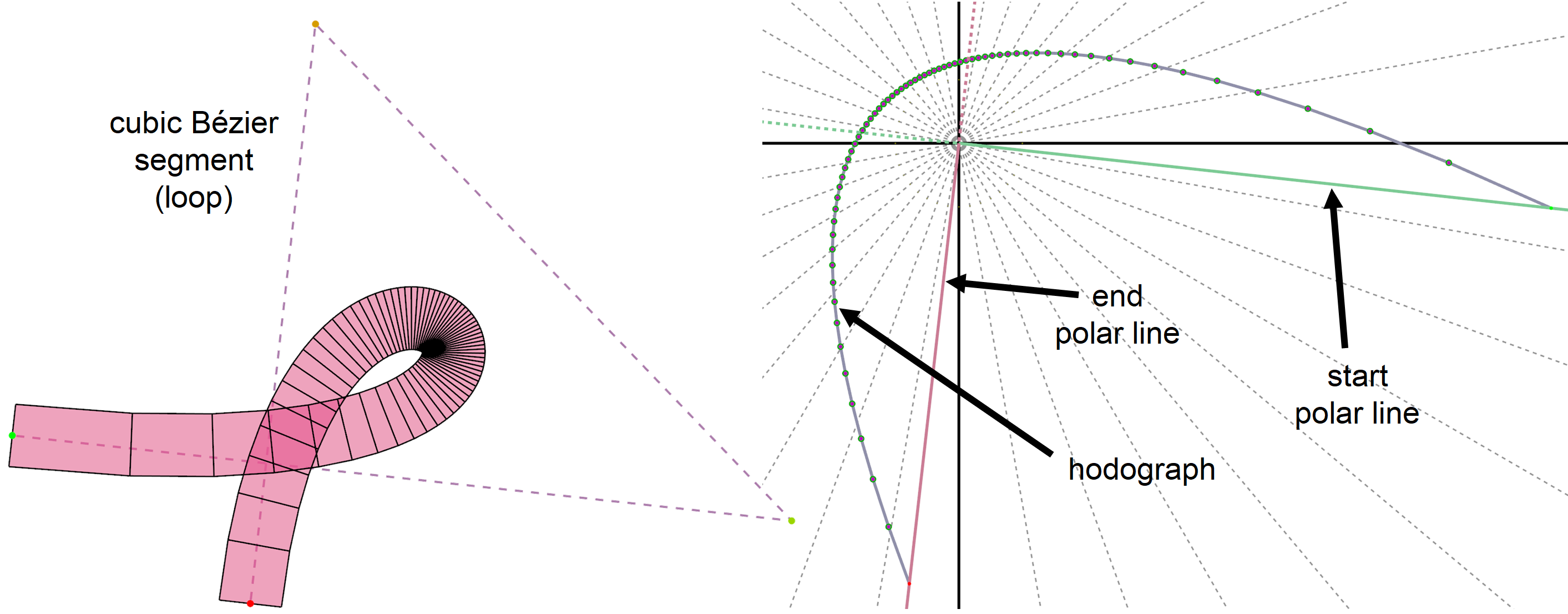}
\caption{\normalfont Cubic \Bezier segment in loop configuration with its hodograph. \label{fig:loop}}
\end{figure}

\begin{figure}
\centering
\includegraphics[width=\columnwidth]{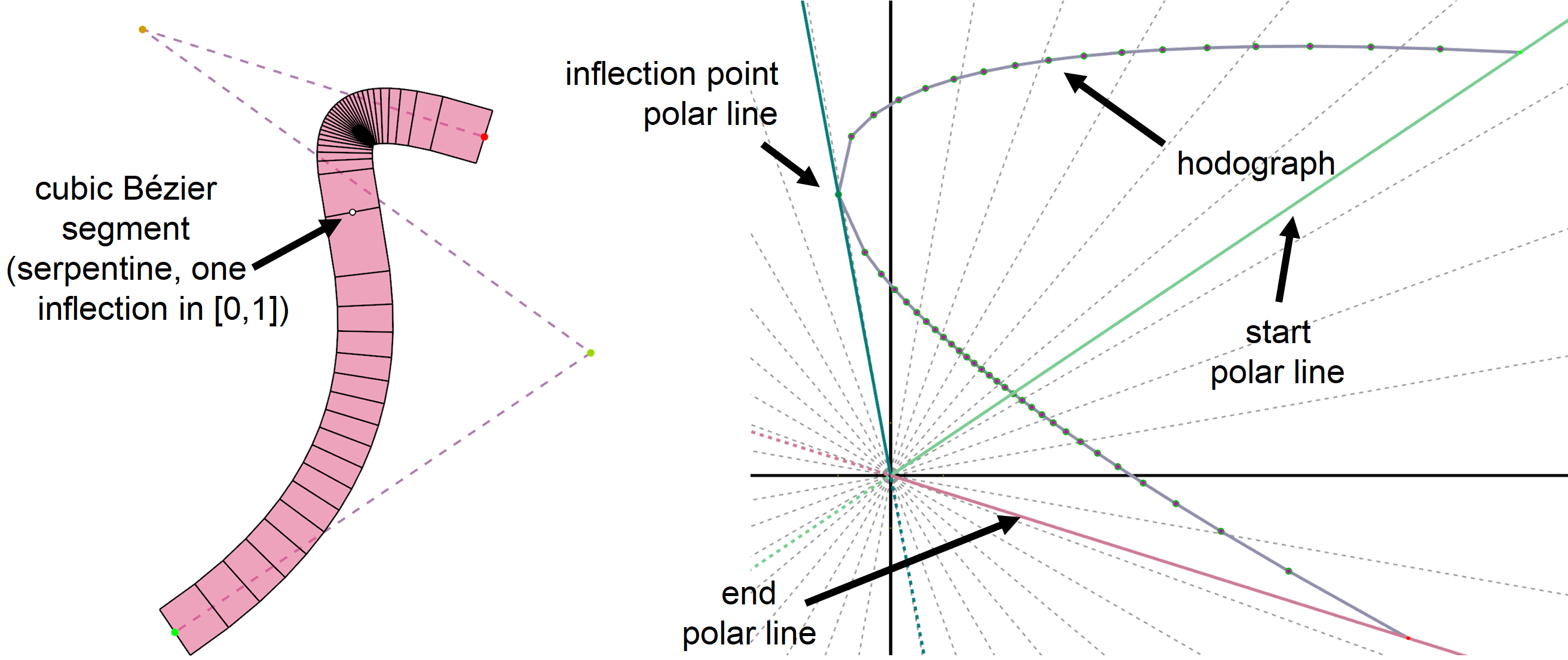}
\caption{\normalfont Cubic \Bezier segment in serpentine configuration (1 inflection in $[0,1]$) with its hodograph. \label{fig:serpentine}}
\end{figure}

\begin{figure}
\centering
\includegraphics[width=\columnwidth]{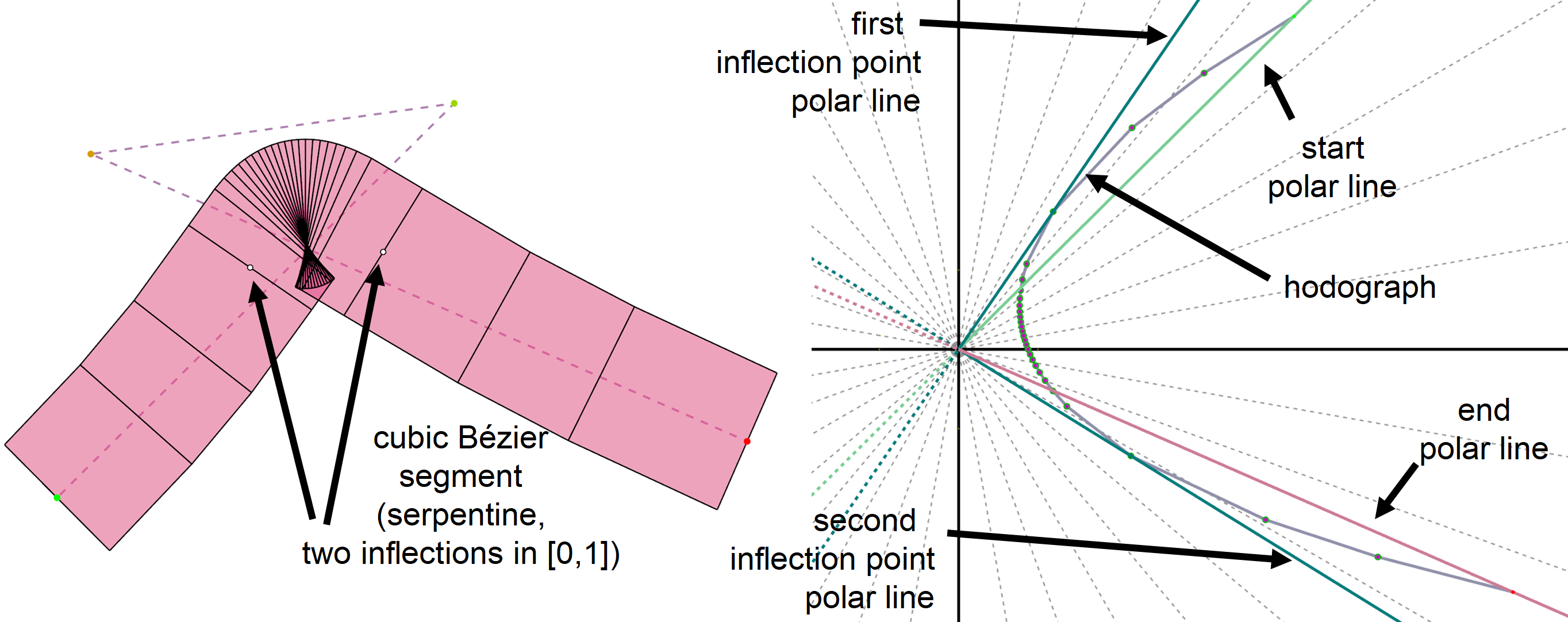}
\caption{\normalfont Cubic \Bezier segment in serpentine configuration (2 inflections in $[0,1]$) with its hodograph. \label{fig:serpentine_2_inflections}}
\end{figure}

\subsubsection{Handling Cusps Robustly}

Consider the rendering implications of a cusp on $\mathbf{g}$, meaning there is a $t$ where $\mathbf{g'}$ passes through the origin $(0,0)$, a situation that occurs when $||\mathbf{g'}(t)||=0$.  By formulating stroking as $\mathbf{s}_{w}(t,u)$, situations where $\mathbf{g'}(t)$ would nearly---but not exactly---pass through $(0,0)$ should induce the segment's normal to ``pivot'' 180\si{\degree} at the limit.  This ever-so-nearly 180\si{\degree} pivot acts to sweep out pixels in a nearly circular (more accurately: double semicircle) region.  However if a cusp formed {\em exactly}---not simply nearly so---then the normal would instantaneously reverse {\em without a pivot} as $\mathbf{t}$ and hence $\mathbf{n}$ are undefined at a cusp so the stroked region defined by $\mathbf{s}_w$ would lack a double semicircle at the cusp point.

Not forming the rasterization coverage of a double semicircle at an exact cusp is inconsistent with how stroking is well implemented in modern path rendering systems.  Additionally the behavior of $\mathbf{s}_{w}$ is practically and artistically undesirable as it permits small perturbations of control points to ``wink away'' stroked coverage if the perturbations were to induce an exact cusp.

We can ``fix'' this undesired behavior of $\mathbf{s}_{w}$ at exact cusps by augmenting the stroked region with a double semicircle of additional points around exact cusps.  See the exact cusp blue region in Figure~\ref{fig:stroked-region}.  We define this augmented stroked region as
\begin{equation}
\label{eq:ConventionalStroking}
\mathbf{S}_{w}(t,u,v) =
\begin{cases}
 \mathbf{g}(t) \pm \frac{w}{2} \, u \, \langle \cos \, \theta(t,v), \sin \, \theta(t,v) \rangle, \text{ if } \substack{ \lVert \mathbf{g}'(t) \rVert = 0 \; \land \\ 0<t<1 } \\
 \mathbf{g}(t) \pm \frac{w}{2} \, u \, \mathbf{n}(t), \text{ otherwise }
\end{cases}
\end{equation}
for $t,u,v \in [0,1]$, where
\begin{align*} 
\theta(t,v) &= \theta_{in}(t) + v\,(\theta_{out}(t) \ominus \theta_{in}(t)) \\
\theta_{in}(t) &= \lim_{s \to t-} \tan^{-1} \mathbf{n}(s) \\
\theta_{out}(t) &= \lim_{s \to t+} \tan^{-1} \mathbf{n}(s) \\
\theta_1 \ominus \theta_2 &=
\begin{cases}
\theta_1 - \theta_2 - 2\pi, & \text{ if } \theta_1 - \theta_2 > +\pi \\
\theta_1 - \theta_2 + 2\pi, & \text{ if } \theta_1 - \theta_2 < -\pi \\
\theta_1 - \theta_2, & \text { otherwise }
\end{cases}
\end{align*}
understanding that $\tan^{-1}$ returns the angle of a vector and $\ominus$ is a relative angle difference.
Notice $\theta(t,v)$ is constructed such that
\begin{align*} 
\theta(t,0) &= \theta_{in}(t)
\\
\theta(t,1) &= \theta_{out}(t)
\end{align*}
If $\mathbf{g'}$ passes through the origin at $u$, $| \theta_{out}(u) \ominus \theta_{in}(u) |$ is 180\si{\degree} as $\mathbf{g}$ is smooth.  Examine the origin of the cusp hodograph in Figure~\ref{fig:cusp} to see this.  The careful formulation above using one-side limits will prove useful to handle caps and joins.  Robustly handling cusps is important not merely to handle segments containing cusps but also to generalize our theory to caps and joins by treating them as essentially partial cusps; see Section~\ref{sec:joins-and-caps} and Figures~\ref{fig:joins} and \ref{fig:caps}.

When the generator curve $\mathbf{g}$ does contain an exact cusp within its parametric domain, $\mathbf{S}_{w}$ unions in a double semicircle pivot at the cusp into the stroked region.
Effectively at a cusp on $\mathbf{g}$, $\mathbf{S}_{w}$ selects an {\em alternative} guaranteed-to-be-defined normal that varies with $\theta(v)$ rather than relying on $t$ varying and its undefined-at-cusps $\mathbf{n}(t)$ normal.

\subsection{Defining a Gradient-Free Stroked Region}

The $\mathbf{S}_{w}$ formulation of stroking in Equation~\ref{eq:ConventionalStroking} allows determining a stroked region without evaluating the generator curve's gradient $\mathbf{g'}$ at cusps.  We now take this idea one step further to forgo depending on the evaluation and normalization of $\mathbf{g'}$ at all.

Rather than varying $t$ over the generator curve $\mathbf{g}$, we {\em instead} explore varying the normal angle $\theta$  by assuming there exists a function $t(\theta)$ that maps a normal angle to a parametric value $t$ that we then use to evaluate $\mathbf{g}$.  This assumed function $t(\theta)$ behaves like the inverse of actual-function $\theta(t)$ defined as
\[
\theta(t) = \tan^{-1} \left( \frac{\mathbf{g'}}{||\mathbf{g'}||} \right) + \frac{\pi}{2}
\]
but with the caveat that $t(\theta)$ is defined even when $||\mathbf{g'}||=0$ and hence even when $\mathbf{n}$ is undefined.

Indeed when there exists some $\theta_c$ such that $t(\theta_c)$ returns a particular value $t_c$ that locates a cusp on $\mathbf{g}$ (so $||\mathbf{g'}(t_c)||=0$), then either $\theta_c$ is among a 180\si{\degree} range of other normal angles that {\em all} return the cusp location $t_c$ (this we call the {\em ordinary cusp} case) or with extreme rarity $\theta_c$ is an isolated angle such as occurs when {\em both} $||\mathbf{g'}(t_c)||=0$ {\em and} $||\mathbf{g''}(t_c)||=0$.  This latter rare case occurs when $\mathbf{g'}$ ``kisses'' the origin but then reverses direction without passing through the origin, a feature known as a {\em kink} \cite{porteous1994geometric}.

Of the four parametric equation forms used in path rendering, only $\mathbf{g}_C$ can form nondegenerate cusps and only a degenerate form of $\mathbf{g}_C$ can form a kink and then only when the degenerate cubic segment is masquerading as a line segment.

Assuming $t(\theta)$ is a one-to-one function over some restricted domain $[\theta_a,\theta_b]$, we can construct a {\em gradient-free} formulation of the stroked region defined by
\[
\mathcal{S}_w(\theta,u) = \mathbf{g}(t(\theta)) \pm \frac{w}{2} \, u \, \langle \cos \, \theta, \sin \, \theta \rangle
 \; \text{ for } \substack{u \in [0,1], \\ \theta \in [\theta_a,\theta_b]}
\]
where $\theta_a$ and $\theta_b$ designate the start and stop normal angles of the stroked region.

\subsubsection{Switching from Normal Angle to Tangent Angle}

So far, this discussion has used normal angles, but expressing $\mathcal{S}_w$ in terms of tangent angles instead will prove more convenient.
Every normal angle $\theta$ is related to its tangent angle $\psi$ by a 90\si{\degree} rotation:
\[
\psi = \theta -\frac{\pi}{2}
\]
Rewriting $\mathcal{S}_w$ in terms of $\psi$ gives
\[
\mathcal{S}_w(\psi,u) = \mathbf{g}(t(\psi)) \pm \frac{w}{2} \, u \, \langle -\sin \, \psi, \cos \, \psi \rangle
 \; \text{ for } \substack{u \in [0,1], \\ \psi \in [\psi_a,\psi_b]}
\]

\subsubsection{The Need for Limited Tangent Angle Ranges}
\label{sec:angle-ranges}

We need $\psi(t)$ to be one-to-one within a bounded range of $t$ so we can invert it to obtain a well-defined function $t(\psi)$ over a range $\psi \in [\psi_a,\psi_b]$.

However when $t(\psi)$ is unconstrained in its range, it may be a multifunction.  Multiple points on $\mathbf{g}$ may share the same tangent angle.  Indeed examples
where $t(\psi)$ is a multifunction are easy to identify.  For example, a spiral or periodic function will share a single tangent angle with many distinct points.  In the extreme, a line segment has a
{\em single} tangent angle for {\em every} $t$.

For ranges of $\psi$ free of points with zero curvature on $\mathbf{g}$ (so containing neither a line segment nor being a curved segment containing an inflection point), $t(\psi)$ can be one-to-one.

Inflection points occur when $\kappa(t)=0$.  If the tangent angle increases (decreases) along a curve, when passing through an inflection point, the tangent angle reverses direction and begins decreasing (increasing) as a consequence of the curve's curvature reversing its sign.  As  $\mathbf{g}$ is smooth, this implies the curve must be revisiting tangent angles---and $t(\psi)$ cannot be one-to-one in this interval.

Stated more simply, a first necessary requirement for $t(\psi)$ to be invertible is its domain must be constrained so all the domain's tangent angles strictly rotate either all clockwise or all counterclockwise.
A second necessary requirement is each angular interval must be less than a complete revolution so $|\psi_b - \psi_a| < 2\pi$.

Quadratic \Bezier $\mathbf{g}_Q$ and linear $\mathbf{g}_L$ segment forms need not solve $\kappa(t)=0$ as these forms are free of distinct inflection points.
For the cubic \Bezier $\mathbf{g}_C$ segment form, 
Loop and Blinn~\shortcite{RenderingVectorArtOntheGPU} provide efficient expressions to setup a quadratic equation to solve for the parametric value $t$ at 0, 1, or 2 inflection points, corresponding to loop, cusp, or serpentine cubic curve types respectively.

\subsubsection{Handling External Conics}

Extra care must be taken for the conic $\mathbf{g}_K$ segment form because we allow negative values of $w_B$.  Non-degenerate conic segments are free of regions where $\kappa(t)=0$.  However particular conic sections we call discontinuous (or external) hyperbolic or parabolic segments have tangent angle reversals when we allow $w_B \leq -1$.  So we use $\mathcal{K}(t)=0$ as a more technical definition for when a tangent angle reversal occurs, defined as
\[
\mathcal{K}(t) = \sgn \lim_{s \to t+} \kappa(s) + \sgn \lim_{s \to t-} \kappa(s)
\]
meaning the signs of the curvature are opposite on either side of $t$ at the limit, or informally the curvature's sign flips moving through $t$.  Note for the (nonrational) forms $\mathbf{g}_C$, $\mathbf{g}_Q$, and $\mathbf{g}_L$, $\mathcal{K}(t) = \kappa(t)$.

When $\kappa_K$ is the curvature of $\mathbf{g}_K$, the numerator of $\kappa_K$ is an involved expression but nonzero---except if $\mathbf{g}_K$ is degenerate, such as flattened to a line segment or point.  Yet the denominator of $\kappa_K$ is much simpler:
\begin{align*}
\denom \kappa_K &= ((1-t)^2 \, + 2 (1-t) t \, w_{B}  + t^2)^3 \\
                          &= ( \denom \mathbf{g}_K )^3
\end{align*}
Notice the denominator of $\kappa_K$ is the cube of the denominator of $\mathbf{g}_K$ (see Conic row of Table~\ref{tab:path-segments}).  As the denominator is smooth, we can solve for when $\denom \mathcal{K}(t)=0$ to
know when its (infinite) curvature reverses.  So the solutions $t_{\text{rev}}$ when $\mathcal{K}_K(t) = 0$ for nondegenerate conic segments $\mathbf{g}_K$ are
\begin{equation}
\label{eq:rev-t}
t_{\text{rev}} = \frac{-2 \pm 2 \sqrt{w_B^2 -1}}{4 w_B -4}
\end{equation}
Our interest is only in solutions in the parametric range $[0,1]$.  There are two solutions when $w_B<-1$, the case of an external hyperbola; one solution when $w_B=-1$, an external parabola; and no solutions (so no tangent reversals) for external ellipses ($-1$ < $w_B < 0$), degenerate lines ($w_B=0$), internal ellipses ($0 < w_B < 1$), internal parabolas ($w_B=1$), or internal hyperbolas ($w_B>1$).

\subsubsection{Building Tangent Angle Ranges of Consistent Turning}

The second requirement for $t(\psi)$ to be one-to-one is its total tangent angle domain must turn less than $2\pi$ radians; otherwise a single tangent angle aliases to more than one parametric value $t$.

By solving $\mathcal{K}(t)=0$ for $t$ strictly inside $[0,1]$, we can produce an ordered sequence $p_{[0...n]}$ of $n+1$ parametric values, defined as
\begin{align}
p_0 &= 0
\nonumber
\\
p_i &= t \, \text{ such that } \, \substack { \displaystyle \mathcal{K}(t)=0 \; \land \\
                                              \displaystyle t \in (0,1) \; \land  \\
                                              \displaystyle p_{i-1} < t }
\label{eq:p-value}
\\
p_n &= 1
\nonumber
\end{align}
such that $i = 1...n-1 \; (i \neq n$).
When $n>1$, $p_1$ through $p_{n-1}$ each identify a tangent angle reversal on $\mathbf{g}$.  Because of the requirement that $p_{i-1} < p_i$ in the definition of $p$, any region of continuous zero curvature is jointly characterized in $p$ by a single $p_i$ value for $i<n$.

Next we produce a second ordered sequence $\Psi_{[0...n]}$ of $n+1$ tangent angles where $\Psi_i$ corresponds to inflections point $\mathbf{g}(p_i)$, defined as
\begin{align}
\Psi_0 &= \tan^{-1} \, \widehat{\nabla} \, \mathbf{g}(0)
\nonumber
\\
\Psi_i &= \tan^{-1} \, \mathbf{g'}(p_i)
\label{eq:Psi}
\\
\Psi_n &= \tan^{-1} \, \widehat{\nabla} \, \mathbf{g}(1)
\nonumber
\end{align}
where $i = 1...n-1 \; (i \neq n)$ and
$\widehat{\nabla}$ is a special normalized gradient operator (detailed shortly) guaranteed in-almost-all-cases to return a well-defined tangent at the start or end point of a parametric curve $\mathbf{g}$, even when $||\mathbf{g'}(0)||=0$ or $||\mathbf{g'}(1)||=0$ respectively.  The one exception to the guarantee is if the segment has an arc length of zero, but then the segment's stroked region is the empty set.

When $i>1$, 
$\Psi_1$ through $\Psi_{n-1}$ are the points of tangent angle reversals $\mathbf{g}(p_1)$ through $\mathbf{g}(p_{n-1})$.  These tangent angles are well-defined because $\mathbf{g'}(p_i)$ for $i \in [1...n-1]$ is well-defined since $\mathcal{K}(p_i)=0$ implies $\mathbf{g'}(p_i)$ exists.

We define $\widehat{\nabla} \, \mathbf{g}(0)$ and $\widehat{\nabla} \, \mathbf{g}(1)$ as
\begin{align*}
\widehat{\nabla} \, \mathbf{g}(0) &= \frac{\lim_{t \to 0+} \mathbf{g'}(t)}{||\lim_{t \to 0+} \mathbf{g'}(t)||} \\
\widehat{\nabla} \, \mathbf{g}(1) &= \frac{\lim_{t \to 1-} \mathbf{g'}(t)}{||\lim_{t \to 1-} \mathbf{g'}(t)||}
\end{align*}
Table~\ref{tab:path-segments} provides $\widehat{\nabla} \, \mathbf{g}(0)$ and $\widehat{\nabla} \, \mathbf{g}(1)$ for each of the four generator function forms.

These definitions rely on the initial and terminal tangent property of the \Bezier basis.  Successive control points are differenced until a nonzero length vector difference is found---or the segment's control point sequence is exhausted.  Using these normalized gradient operators, the $\Psi_0$ and $\Psi_n$ tangent angles are well-defined for nonzero length segments, even when one or more control points---but not all---are colocated.  All control points being colocated is a zero length segment.

\subsubsection{Bounding Total Curvature Within Tangent Angle Intervals}

We now consider the possibility that $\mathcal{S}_w(\psi,u)$ might not be a one-to-one function in one or more of the intervals $[\Psi_i,\Psi_{i+1}]$ because the tangent angle ``wraps around'' a full turn (i.e., $2\pi$ radians) or more.  We know there are cases such as if $\mathbf{g}$ is a spiral when we can expect total curvature to exceed $2\pi$.  However we limit our consideration to just the four parametric equation forms defined for path segments in Table~\ref{tab:path-segments} and Section~\ref{sec:path-math-prelim}.

Rational \Bezier curves with nonnegative homogeneous weights adhere to the {\em hodograph property} \cite{floater1992derivatives}.
This property says the segment's tangent (in the direction of increasing $t$) lies between the directions of the control polygon segments $\mathbf{P}_{i+1}-\mathbf{P}_i$.

So for the quadratic $\mathbf{g}_Q$ form with 3 control points,
all the segment's tangents must be between $\mathbf{P}_1-\mathbf{P}_0$ and $\mathbf{P}_2-\mathbf{P}_1$.  
The maximum angle between such a pair of segments is $\pi$ radians.
Therefore a quadratic \Bezier path segment $\mathbf{g}_K$ has a maximum absolute total angle range of $\pi$.
For the cubic  $\mathbf{g}_C$ form with 4 control points, all the segment's tangents must be between two pairs of such segments.
Therefore a cubic \Bezier path segment $\mathbf{g}_C$ has a maximum absolute total angle range of $2\pi$.
The maximum total angle range of the linear $\mathbf{g}_L$ form is trivially zero as a line is straight so has no tangent angle change.

The conic $\mathbf{g}_K$ form deserves more discussion.
$\mathbf{g}_K$ is a rational quadratic \Bezier curve where we allow the weight $w_B$ to be either nonnegative {\em or} negative.
When $w_B \geq 0$, the weights are all nonnegative, satisfying the conventional hodograph property, so the maximum total angle change is $\pi$ just like for $\mathbf{g}_Q$.
However when $-1 \leq w_B<0$, $\mathbf{g}_K$ can ``flex outward'' so its tangent angle range is the reflex of directions that lie between $\mathbf{P}_1-\mathbf{P}_0$ and $\mathbf{P}_2-\mathbf{P}_1$ so the absolute angle of $\mathbf{g}_K$ with a weight $-1 < w_B < 0$ would be between $\pi$ and $2\pi$ radians.  Finally when $w_B < -1$ the gradient direction range is bounded between the limit of the gradient direction of $t_{\text{rev}}$ from Equation~\ref{eq:rev-t} so here the maximum total angle change is $\pi$.

Based on this analysis, for {\em all} the parametric equation forms 
that path rendering standards use, the total curvature of any interval $[\Psi_i,\Psi_{i+1}]$ is $< 2\pi$.  This means there is no need to split $[\Psi_i,\Psi_{i+1}]$ intervals to be $< 2 \pi$.  The sequences $p$ and $\Psi$ are limited to a maximum of 4 elements because they need 2 elements for the initial and terminal elements for $t=0$ and $t=1$ and at most 2 more elements for the at most 2 inflection points allowed by the $\mathbf{g}_C$ or $\mathbf{g}_K$ forms.  With at most 4 elements in the sequence $\Psi$, there are at most 3 intervals.

In the case of a conic path segment when $-1 < w_B < 0$ (external ellipse) or a cubic \Bezier segment without multiple inflections (so a loop or cusp cubic), a single interval could have a total turning angle $\geq \pi$.  In this situation, we find it numerically helpful (see Section~\ref{sec:Solving_for_t}) to split the region into two intervals $[\Psi_0,\text{split}(\Psi_0,\Psi_1)]$ and $[\text{split}(\Psi_0,\Psi_1),\Psi_1]$ where
$\text{split}(\theta_a,\theta_b)$ is defined
\[
\text{split}(\theta_a,\theta_b) = \theta_a \oplus \frac{\theta_b \ominus \theta_a}{2} \oplus \pi
\]
and $\oplus$ is angle addition defined as
\begin{align*}
\theta_1 \oplus \theta_2 =
\begin{cases}
\theta_1 + \theta_2 - 2 \pi, & \; \text{ if } \, \theta_1 + \theta_2 > +\pi \\
\theta_1 + \theta_2 + 2 \pi, & \; \text{ if } \, \theta_1 + \theta_2 < -\pi \\
\theta_1 + \theta_2, & \; \text{ otherwise }
\end{cases}
\end{align*}
Splitting such intervals in half so each half has $<\pi$ radians makes it numerically unambiguous to distinguish an angle $\psi$ from $\psi+\pi$ in the process of evaluating $t(\psi)$.

\subsubsection{Building a Unified Tangent Angle Interval Range}
\label{sec:UnifiedTangentAngleIntervalRange}

After establishing our intervals as described, we have 1 to 3 intervals---call this the interval count $M$.  Each interval's $t(\psi)$ is one-to-one, except in the case of a flat interval where $\Psi_{i-1} = \Psi_i$, such as a line segment $\mathbf{g}_L$ form or a degenerate version of some other segment form.

We specify the radian difference $\delta_i$ in each interval and the accumulated absolute $\delta\Sigma(k)$ for each interval as
\begin{align}
\delta_i &= \Psi_{i-1} \ominus \Psi_i
\label{eq:continuous-delta}
\\
\delta_\Sigma(k) &= \sum_{i=1}^{k} |\delta_i| \; \text{ for } \, k \in [0...M]
\nonumber
\end{align}
so that $\delta_\Sigma(0) = 0$ and $\delta_\Sigma(M)$ is total absolute angle rotation over all the intervals.
By how we constructed our intervals, we know $\delta_i \leq \pi$ and $\delta_\Sigma \leq 2\pi$ for $\mathbf{g}_f$ where $f \in {C, Q, K, L}$.
So all the standard path segment forms listed in Table~\ref{tab:path-segments} rotate no more than 180\si{\degree} in any interval and no more than 360\si{\degree} total.

For a value $z \in [0,\delta_\Sigma(M)]$, we specify a function $\psi(z)$ as
\begin{align*}
\psi(z) =
\begin{cases}
\Psi_k, & \text{ if } \, \exists \, k : z = \delta_\Sigma(k) \\
\Psi_k + \sgn \delta_k \; (z - \delta_\Sigma(k)), & \text{ else } \, \substack{ \exists \, k : \\ \delta_\Sigma(k) < z < \delta_\Sigma(k+1) }
\end{cases}
\end{align*}
By building on $\psi(z)$, we can define robust functions on $z$ that return a parametric value $t(z)$ and a unit tangent $\mathbf{n}(z)$:
\begin{align}
t(z,v) =
\begin{cases}
p_k, & \text{ if } \, \exists \, k : z = \delta_\Sigma(k)  \land \delta_i \neq 0 \\
t_{[\Psi_k,\Psi_{k+1}]}(\psi(z)), & \text{ else if } \, \substack{ \exists \, k : \\ \delta_\Sigma(k) < z < \delta_\Sigma(k+1) } \\
(1-v) \, p_k + v \, p_{k+1}, & \text{ otherwise } \, \exists \, k : z = \delta_\Sigma(k)
\end{cases}
\label{eq:PolarStroking}
 \\
\mathbf{n}(z,v) =
\begin{cases}
\langle - \sin \Psi_k, \cos \Psi_k \rangle, & \text{ if } \, \exists \, k : z = \delta_\Sigma(k) \\
\langle - \sin \psi(z), \cos \psi(z) \rangle, & \text{ else } \, \substack{ \exists \, k : \\ \delta_\Sigma(k) < z < \delta_\Sigma(k+1) }
\end{cases}
\label{eq:theory_n_z}
\end{align}
The yet-to-be-defined function $t_{[\Psi_k,\Psi_{k+1}]}(\psi)$ maps an angle $\psi$ to a parametric value $t$ within the interval $[\Psi_k,\Psi_{k+1}]$; our next Section~\ref{sec:Solving_for_t} 
explains the construction of this function.

The {\em otherwise} case in Equation~\ref{eq:PolarStroking} operates for zero curvature intervals, using $v$ to fill in a flat interval with a widened line segment.  Much as $S_w(t,u,v)$ in Equation~\ref{eq:ConventionalStroking}
varies $v$ to generate cusps, $S_w(z,u,v)$ instead varies $v$ to generate widened line segments for flat intervals.

Now we express $\mathbf{S}_w$ in terms of these expressions to arrive at a gradient-free formulation of the stroked region of a path segment $\mathbf{g}$
\begin{align}
\mathbf{S}_{w}(z,u,v) = \mathbf{g}(t(z,v)) \pm \frac{w}{2} \, u \, \mathbf{n}(z) \; \text{ for } \substack{ z \in [0,\delta_{\Sigma}(M)], \\ u,v \in [0,1] }
\label{eq:polar-stroking}
\end{align}
This $\mathbf{S}_{w}(z,u,v)$ version of $\mathbf{S}$ is superior to the $\mathbf{S}_w(t,u,v)$ version in Equation~\ref{eq:ConventionalStroking} because the
former is gradient-free
and provides a way to traverse uniformly the path segment in tangent angle by varying $z$ linearly over $[0,\delta_{\Sigma}(M)]$.  This last point is our
{\em big idea} and the basis for our polar stroking method of tessellation.

We call the conventional theory {\em parametric stroking} (Equation~\ref{eq:ConventionalStroking}) because the parametric variable $t$ drives the generation of the stroked region along the generator curve.
We call our new theory {\em polar stroking} (Equation~\ref{eq:PolarStroking}) because the tangent angle $\psi$, expressed as a polar angle, drives the stroked region along
the generator curve.

To complete our theory, we define a support predicate to indicate when a pixel at $(x,y)$ is inside of the stroked segment using $\mathbf{S}_{w}(z,u,v)$ in Equation~\ref{eq:polar-stroking}:
\begin{align}
\mathbf{S}_{w}(x,y) =
\begin{cases}
1, &\; \text{ if } \exists_{z \in [0,\delta_\Sigma(M)], u,v \in [0,1]} : (x,y) = \mathbf{S}_{w}(z,u,v) \\
0, &\; \text{ otherwise}
\end{cases}
\label{eq:polar-stroking-predicate}
\end{align}
The support coverage for an entire path $P$ is the maximum of the support coverage of each path segment in $P$ according to Equation~\ref{eq:polar-stroking-predicate} and that of any joins and caps.
This provides a robust support predicate for stroked paths comparable to the support predicates for filled paths found in Appendix~\ref{sec:fill-theory}.

Path rendering standards are sufficiently concrete about the regions defined by caps and joins (e.g., handling miters, etc.)
that we do not belabor defining the stroked regions including caps and joins in formal terms.

\subsection{From Tangent Angle to Parametric Value}
\label{sec:Solving_for_t}

We must still explain how to implement the assumed function $t_{[\Psi_k,\Psi_{k+1}]}(\psi)$ in Equation~\ref{eq:PolarStroking}.  This involves solving for $t$ when the gradient is orthogonal to the normal vector $\mathbf{N}$ (90\si{\degree} rotated from the tangent angle $\psi$) so
\begin{align}
\label{eq:angle2t}
0 = \mathbf{g'}(t) \cdot \mathbf{N}
\end{align}
where
\begin{align*}
\mathbf{N} = \langle -sin \, \psi, cos \, \psi \rangle
\end{align*}
and then selecting what {\em should be by construction} the single solution in the range $[p_k,p_{k+1}]$.  However if numerically no solution is in the range $[p_k,p_{k+1}]$, evaluate the range
extremes and pick $t$ using
\begin{align*}
t =
\begin{cases}
p_k, &\; \text{ if } |\mathbf{g'}(p_k) \cdot \mathbf{N}| < |\mathbf{g'}(p_{k+1}) \cdot \mathbf{N}| \\
p_{k+1}, &\; \text{ otherwise }
\end{cases}
\end{align*}
For the cubic $\mathbf{g}_C$ and conic $\mathbf{g}_K$ forms, this involves solving a quadratic equation.  The conic $\mathbf{g}_K$ form also needs solve only a quadratic because we can ignore the denominator of $\mathbf{g'}_Q$ when solving Equation~\ref{eq:angle2t}.  For the quadratic \Bezier $\mathbf{g}_Q$, this involves solving a simple linear equation.  The linear $\mathbf{g}_L$ never needs to perform this solve.

Notice at a cusp, $\mathbf{g'}$ will be (0,0) so any angle will satisfy Equation~\ref{eq:angle2t}---though the $t$ for the cusp might not be in the range of interest $[p_k,p_{k+1}]$.


\section{The Polar Stroking Method}
\label{sec:method}

We now turn this theory into a robust discrete tessellation scheme for a complete path.

The algorithm we seek should have these properties:
\begin{itemize}

\item Degenerate path segments, cubic \Bezier path segments with cusps, and all other valid path segment, caps, and joins should approximate the theory in Section~\ref{sec:theory}.

\item Intuitive control of the tessellated quality; this means the facet angles between tessellated quadrilaterals (called {\em quads} henceforth) are guaranteed less than a configurable facet angle threshold while also uniformly distributing the change in tangent angle within an inflection-bounded interval.

\item The number of tessellated quads must also be determined {\em a priori} to tessellation of a given path segment, as opposed to being the result of a recursive process; this is motivated by wanting to map well to GPUs where predictable work creation is necessary as GPUs do not naturally support recursive processes.

\item Unified handling of joins and caps using the same approach as path segments.

\end{itemize}

\subsection{Stroked Paths Tessellated to Quad Strips}
\label{sec:QuadStrip}

The output of our stroke tessellation algorithm should be a sequence of quads, suited for GPU rendering.  For uniformity of processing, if we need to generate a triangle (such as for a miter join), we generate a degenerate quad with two colocated vertices.  These could easily be optimized into triangles.

We expect these sequences of quads to be rasterized by GPUs designed to rasterize triangles.  Standard practice for GPUs is to subdivide quads into two triangles and rasterize each triangle independently and perform attribute interpolation per-triangle.  This is not ideal for our purposes as our quads ``bow-tie'' \cite{Strassmann:1986:HB:15922.15911} (meaning opposite edges intersect) whereas the GPU draws two triangles that overlap.  We also naturally expect bilinear attribute interpolation over the quad so all 4 vertices contribute---per-triangle interpolation is noticeably inferior.
Hormann and Tarini~\shortcite{Hormann:2004:QRP:1058129.1058131} describe proper methods for rendering a quad with two colocated vertices; we implemented proper quad rasterization and interpolation (adapting a geometry shader example found in the Cg Toolkit \cite{CgToolkit} source code) and found this approach remedied the rasterization and interpolation issues attributable to GPUs splitting quads into triangles.

\subsection{Uniform Tangent Angle Step Tessellation}
\label{sec:TessProcess}

We now assume a maximum tangent angle step threshold $q$.  Treat $q$ as an intuitive tessellation quality knob that determines the maximum tangent angle step along the generator curve.

Farouki and Neff~\shortcite{Farouki:1990:APP:87526.87543} explain that an offset curve's tangent $\mathbf{t}_o$ and normal $\mathbf{n}_o$ vectors, at any parametric $t$, are a linear scale factor different from the tangent $\mathbf{t}$ and normal $\mathbf{n}$ of the generator curve at the same $t$.  When $\kappa$ is the generator curve's curvature at $t$:
\[
\mathbf{t}_o = \frac{1 \pm \kappa \frac{w}{2}}{|1 \pm \kappa \frac{w}{2}|}\,\mathbf{t}, \;\; \mathbf{n}_o = \frac{1 \pm \kappa \frac{w}{2}}{|1 \pm \kappa \frac{w}{2}|}\,\mathbf{n} \;\;
\]
So the tangent and normal angles, respectively, of offset and generator curves are equal modulo 180\si{\degree}.  Also if the scale factor is zero, the offset curve cusp forms a cusp (as its gradient is zero) and an angle reversal must occur when traversing that cusp.

Thus the tangent and normal angles on the boundary of the stroke change by the same step in angle as the generator curve's tangent and normal angle---except reversing at offset cusps.  Arguably the stroked tessellation quality is more sensitive to what we call the {\em facet angle}, the angle when one quad connects to the next.  Ordinary facet angles are bounded to $< 2q$ though usually quite close to $q$. 
Consult our supplement \mycite{facet-angle} for details.

Hence the reason $q$ is an effective quality knob is $q$ provides a uniform tangent angle step that then bounds the ordinary facet angle change.  This bound excludes a small number of exceptional facet angles adjacent to offset cusps on the boundary that lack a bound and are typically internal to the tessellation.

\subsubsection{Building a Discrete Interval Range}

\begin{figure}
\centering
\includegraphics[width=0.95\columnwidth]{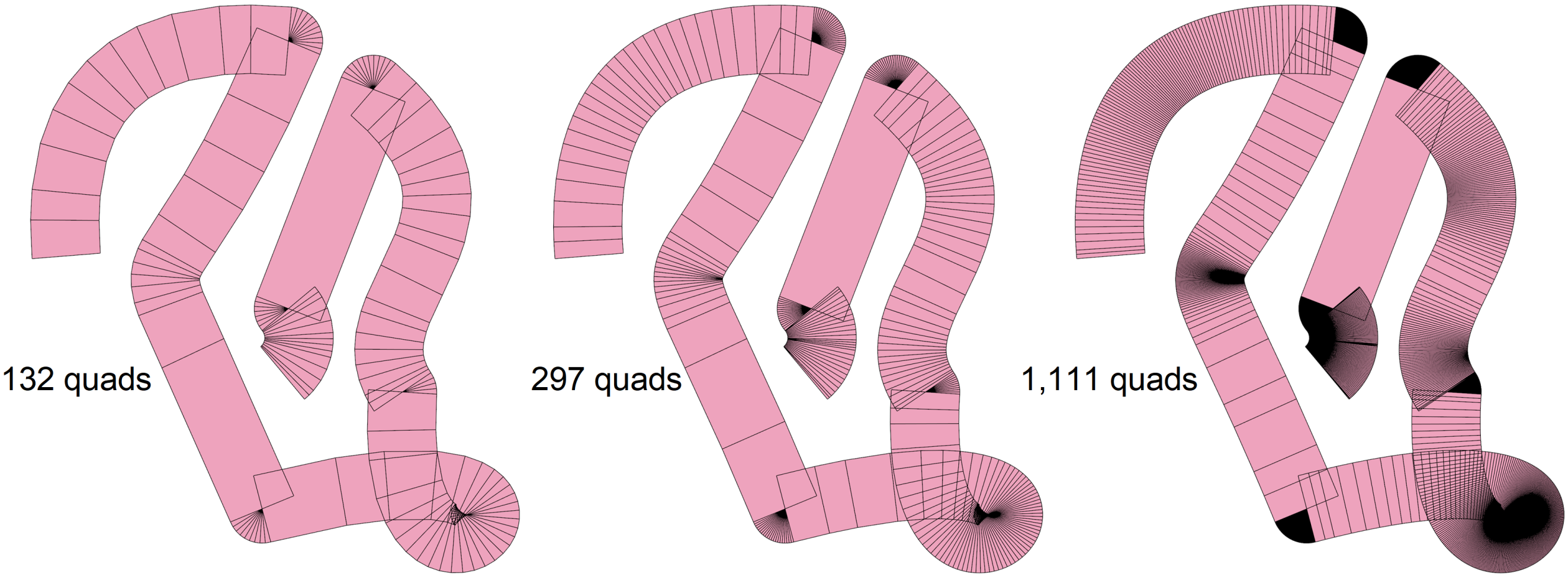}
\caption{\normalfont Path with cubic, quadratic, conic, and linear segments drawn from left-to-right with $q={10\si{\degree}, 4\si{\degree}, 1\si{\degree}}$. \label{fig:varying-max-gradient-angle}}
\end{figure}

To build our tessellation of a path segment, we now compute a number of steps $\Delta_i$ per interval and cumulative number of steps for the segment $\Delta_\Sigma(k)$:
\begin{align}
\Delta_i = \ceil*{ \frac{ | \delta_i |}{q} }
\label{eq:Delta_i}
\\
\Delta_\Sigma(k) = \sum_{i=1}^{k} \Delta_i
\label{eq:DeltaSigmaK}
\end{align}
so that $\Delta_\Sigma(0)=0$ and $\Delta_\Sigma(M) = N$ where $N$ is the total number of steps for the entire path segment.  Figure~\ref{fig:varying-max-gradient-angle} shows how decreasing $q$ affects the tessellation.

For a value $j=0...N$, we can now determine a function $\psi(j)$ such that as $j$ varies from $0$ to $N$, the function steps in $t$ such that the change in absolute tangent angle from $\psi(j)$ to $\psi(j+1)$ is guaranteed to change by $\leq q$.
\begin{align*}
\psi(j) =
\begin{cases}
\Psi_k, & \text{ if } \, \exists \, k : j = \Delta_\Sigma(k) \\
\Psi_k + \frac{ \delta_k}{\Delta_k} \, (j - \Delta_\Sigma(k)), & \text{ else } \, \exists \, k : \Delta_\Sigma(k) < j < \Delta_\Sigma(k+1)
\end{cases}
\end{align*}
By building on $\psi(j)$, we can define robust functions to return a parametric value and unit tangent from stepping in $j$:
\begin{align}
t(j) &=
\label{eq:func_t}
\begin{cases}
p_k, & \text{ if } \, \exists \, k : j = \Delta_\Sigma(k) \\
t_{[\Psi_k,\Psi_{k+1}]}(\psi(j)), & \text{ else } \, \exists \, k : \Delta_\Sigma(k) < j < \Delta_\Sigma(k+1)
\end{cases}
\\
\mathbf{n}(j) &=
\begin{cases}
\langle - \sin \Psi_k, \cos \Psi_k \rangle, & \text{ if } \, \exists \, k : j = \Delta_\Sigma(k) \\
\langle - \sin \psi(j), \cos \psi(j) \rangle, & \text{ else } \, \exists \, k : \Delta_\Sigma(k) < j < \Delta_\Sigma(k+1)
\end{cases}
\label{eq:func_N}
\end{align}


\begin{figure}
\centering
\includegraphics[width=\columnwidth]{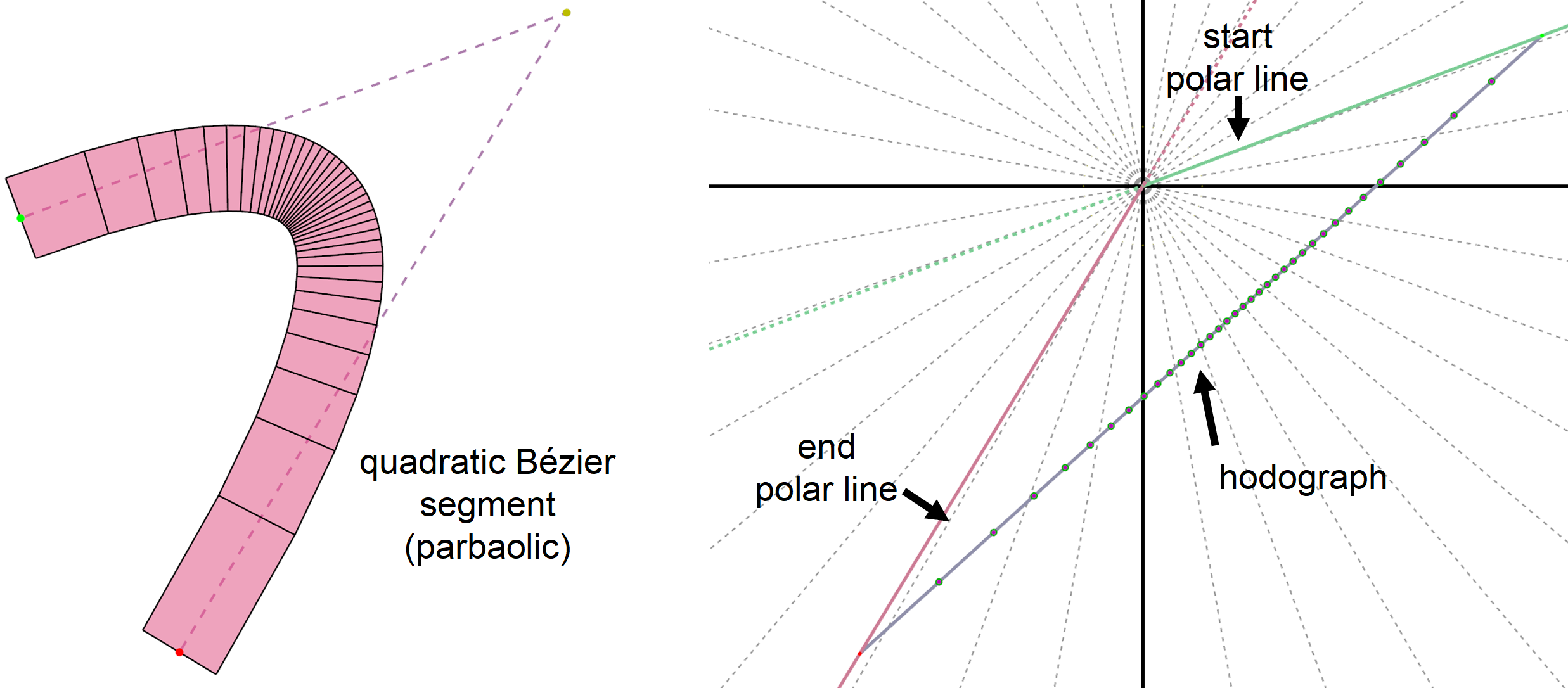}
\caption{\normalfont Quadratic \Bezier segment with its hodograph. \label{fig:quadratic}}
\end{figure}

The function $t(j)$ checks if $j$ corresponds to an interval boundary $\Psi_k$ for some $k$ and, if so, simply
returns the parametric values $p_k$; otherwise, $j$ falls within an interval $[\Psi_k,\Psi_{k+1}]$ and then linearly interpolates a tangent angle $\psi$ in the range to use to evaluate the function $t_{[\Psi_k,\Psi_{k+1}]}(\psi)$ for the interpolated angle.  Likewise $\mathbf{n}(j)$ operates similarly but returns a unit normal corresponding to $t(j)$.

Unlike Equations~\ref{eq:ConventionalStroking} and \ref{eq:theory_n_z} that need a varying $v$ to generate flat segments, discrete tessellation has no such need.  A flat segment will be rasterized as a quad so there is no need for $v$ to generate points.  This means a line segment tessellates to a single quad.

\subsubsection{Tessellating a Path Segment to Quads}

To tessellate a path segment with a particular generator path segment equation $\mathbf{g}_f$, associated control points, and stroke width $w$, first compute $N$ and the sequences $p$, $\Psi$, $\delta$, $\Delta_\Sigma$.

Break the tessellation of a path segment into $N = \Delta_\Sigma(M)$ steps.  $N+1$ ribs are generated, each having a pair of vertices $P_i$ and $N_i$ where $j = 0...N$ defined
\begin{align}
\mathbf{N}_j = \mathbf{g}(t(j)) - r_N \, \mathbf{n}(j)
\label{eq:RibVertices-N}
\\
\mathbf{P}_j = \mathbf{g}(t(j)) + r_P \, \mathbf{n}(j)
\label{eq:RibVertices-P}
\end{align}
where
\[
r_N = \frac{w}{2}, \;\;\; r_P = \frac{w}{2}
\]
The distinct positive- and negative-directed radii $r_N$ and $r_P$ are introduced to aid in generating caps and joins in the next section.

Then generate the $N$ tessellated quads numbered $i=0...N-1$ assembled from pairs of sequential ribs where each has 4 vertices: $\mathbf{N}_i$, $\mathbf{P}_i$, $\mathbf{N}_{i+1}$, and $\mathbf{P}_{i+1}$.


\begin{figure}
\centering
\includegraphics[width=3.3in]{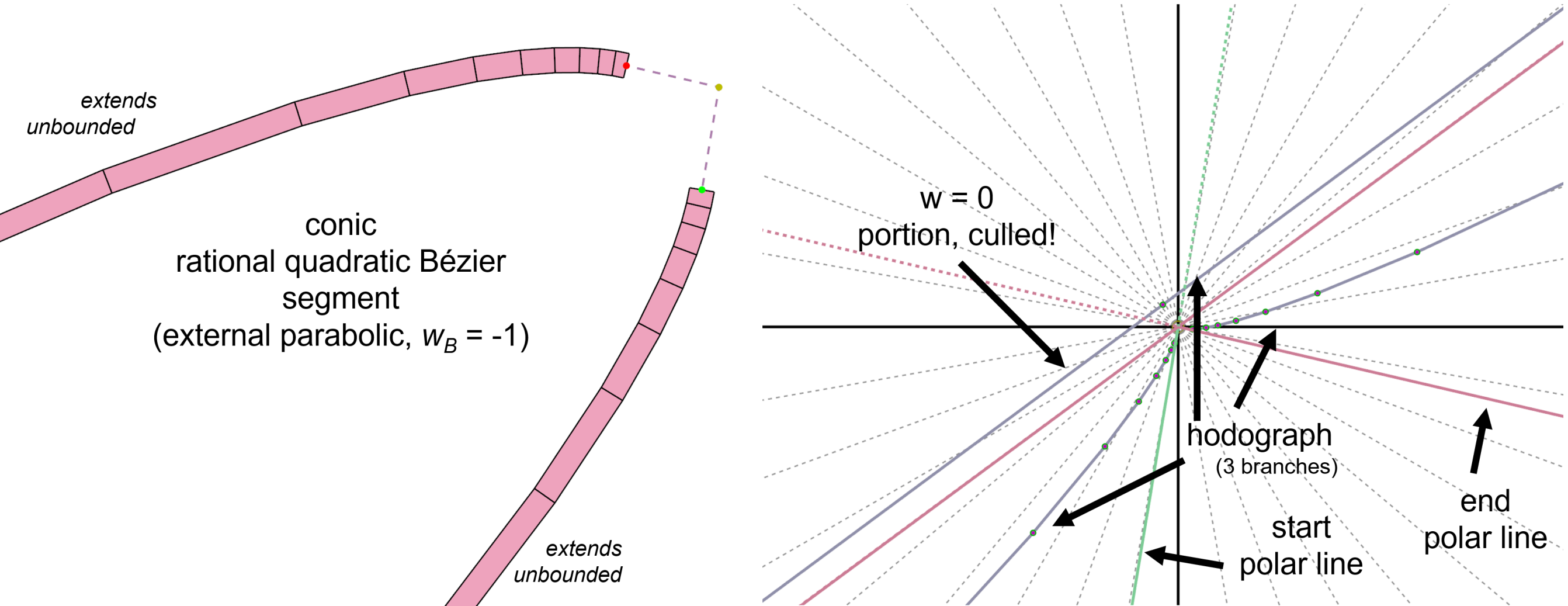}
\caption{\normalfont Rational quadratic \Bezier segment in external parabola configuration with its hodograph. \label{fig:external_parabola}}
\end{figure}

\begin{figure}
\centering
\includegraphics[width=3.1in]{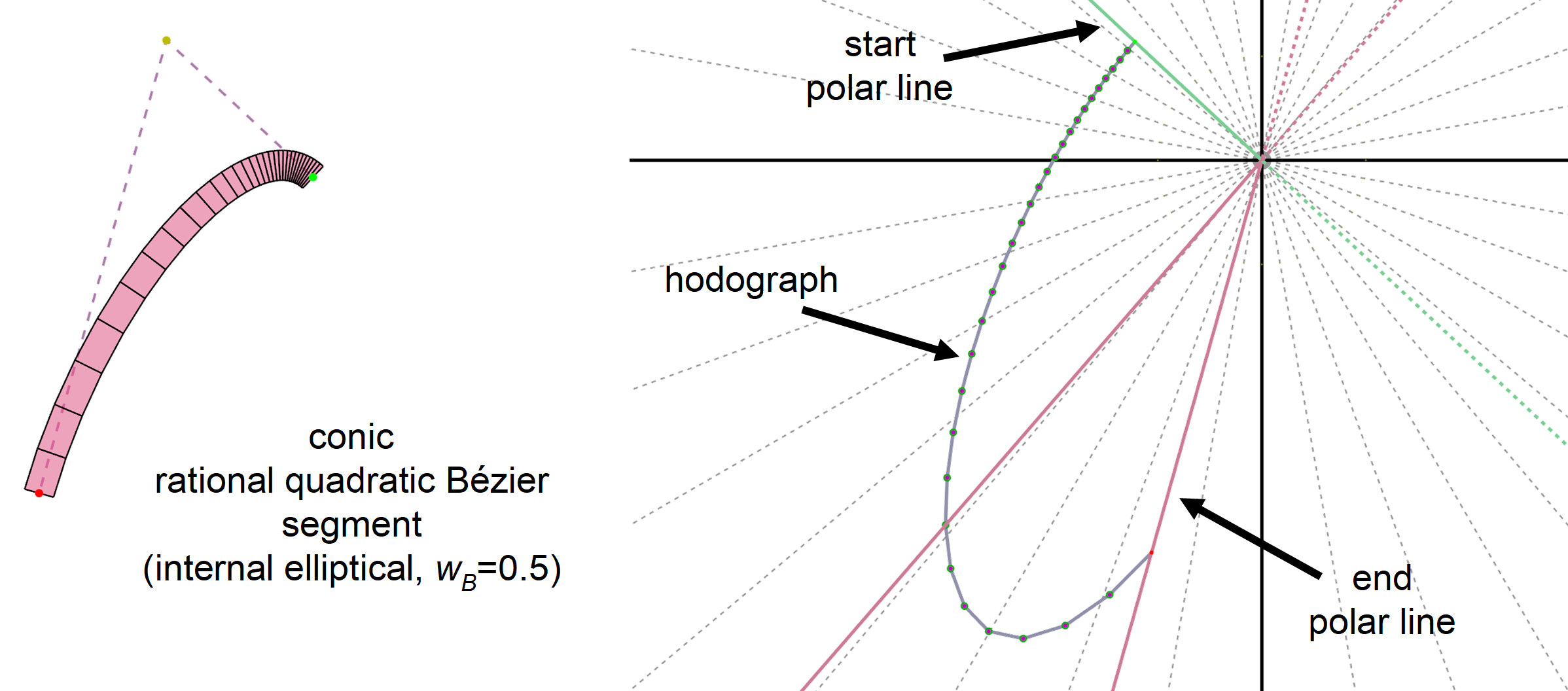}
\caption{\normalfont Rational quadratic \Bezier segment in internal ellipse configuration with its hodograph. \label{fig:internal_ellipse}}
\end{figure}

\begin{figure}
\centering
\includegraphics[width=3.3in]{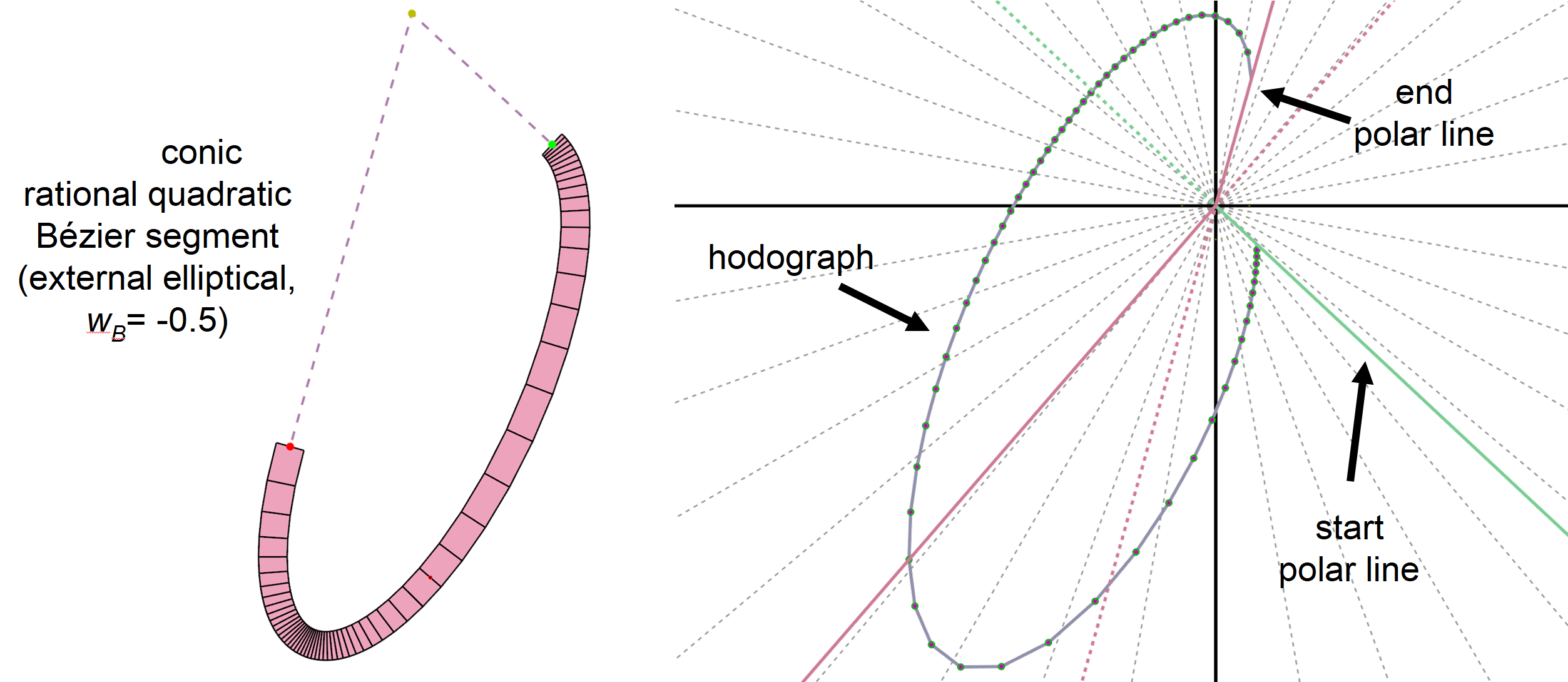}
\caption{\normalfont Rational quadratic \Bezier segment in external ellipse configuration with its hodograph. \label{fig:external_ellipse}}
\end{figure}

\begin{figure}
\centering
\includegraphics[width=3.3in]{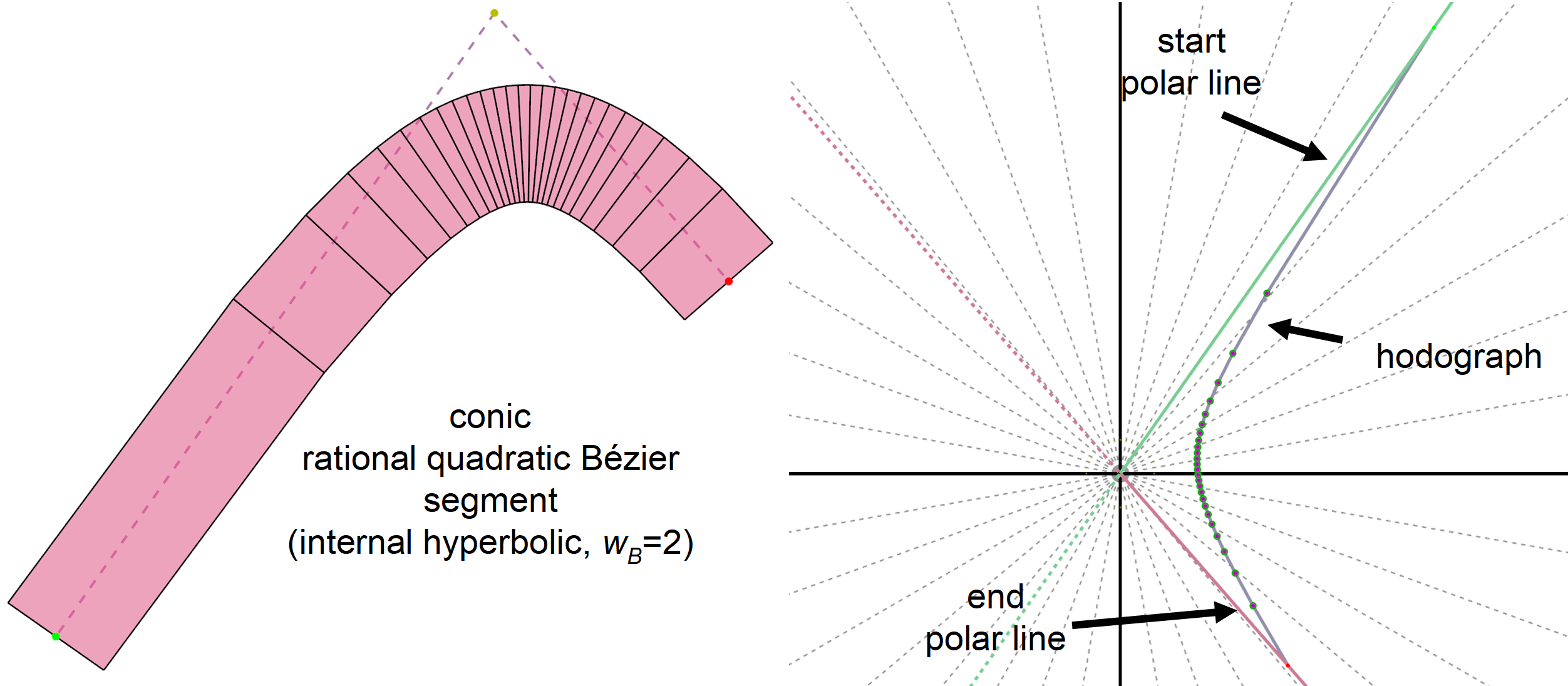}
\caption{\normalfont Rational quadratic \Bezier segment in internal hyperbola configuration with its hodograph. \label{fig:internal_hyperbola}}
\end{figure}

\begin{figure}
\centering
\includegraphics[width=3.3in]{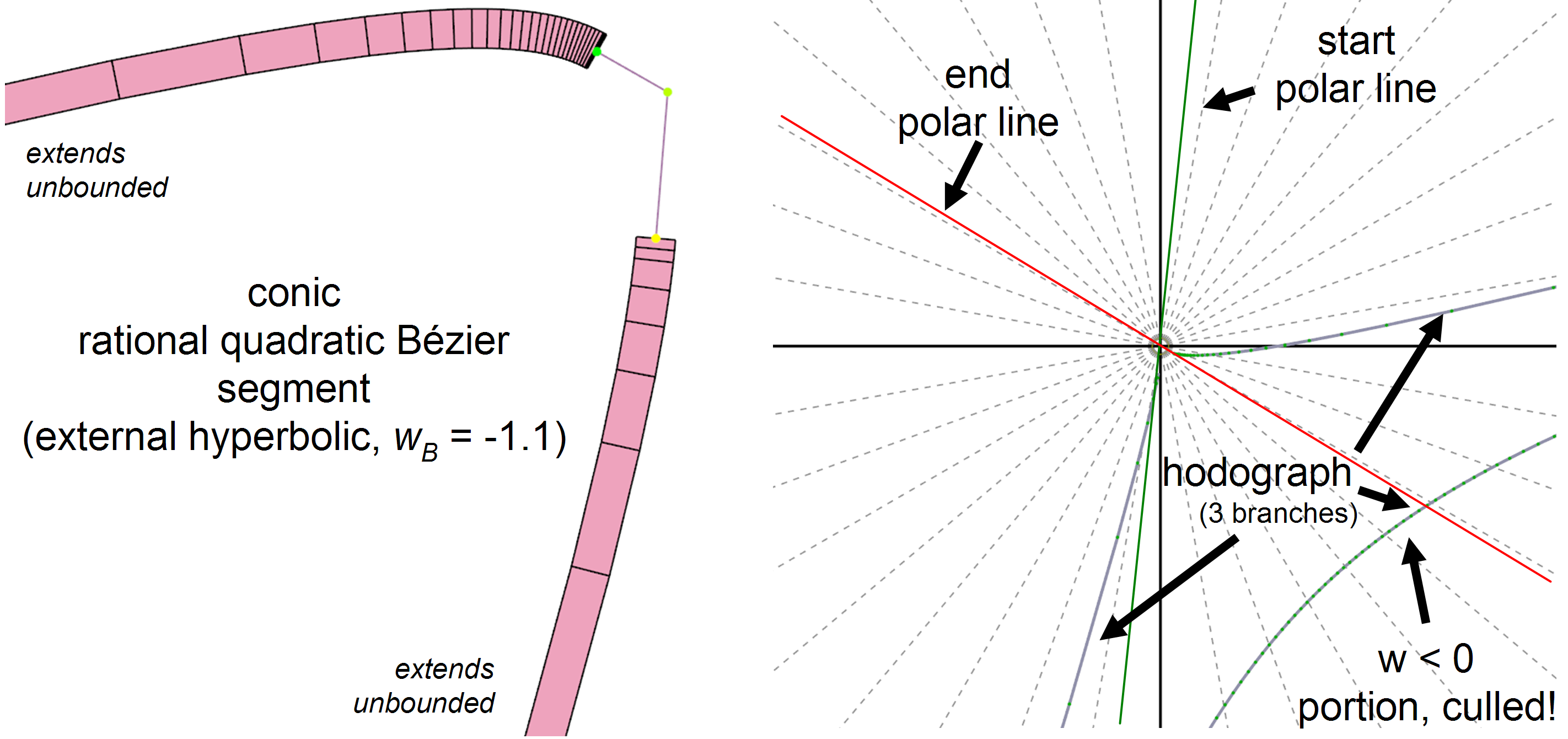}
\caption{\normalfont Rational quadratic \Bezier segment in external hyperbola configuration with its hodograph. \label{fig:external_hyperbola}}
\end{figure}

\subsubsection{Hodograph Intuition}

\begin{figure}
\centering
\includegraphics[width=0.95\columnwidth]{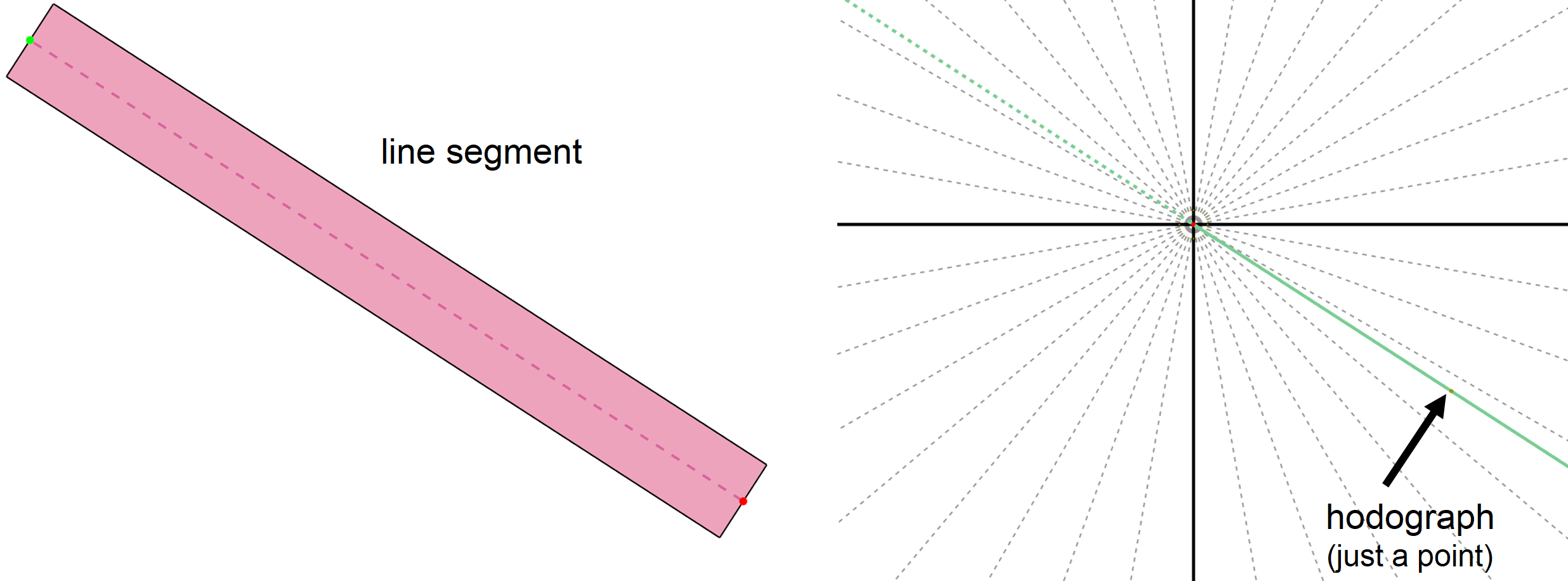}
\caption{\normalfont Linear segment with its hodograph. \label{fig:line_segment}}
\end{figure}

To help appreciate our approach, Figures~\ref{fig:loop} to \ref{fig:line_segment} 
show on their right side a path segment, tessellated by our GPU-based implementation of polar stroking and overlaid with its wireframe tessellation while the right side shows the hodograph (a polar plot of the gradient $\mathbf{g'}$) of the segment.  Points on the hodograph correspond to ribs on the tessellated path segment.

\subsection{Joins and Caps}
\label{sec:joins-and-caps}

\begin{figure}
\centering
\includegraphics[width=0.95\columnwidth]{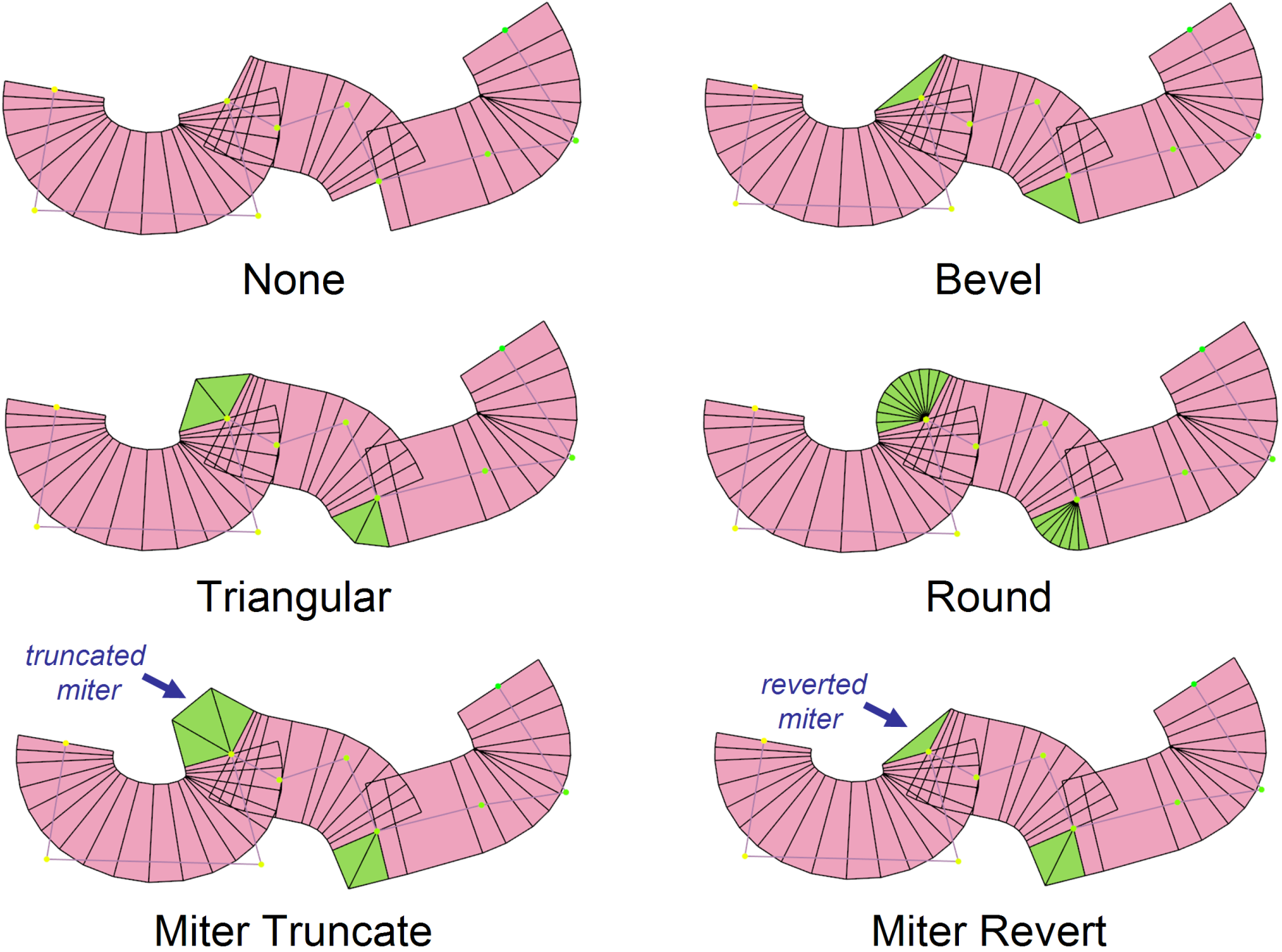}
\caption{\normalfont Tessellations of all supported join styles. \label{fig:joins}}
\end{figure}

When consecutive path segments share the same end and start points but do not join with exact tangent continuity---as is often the case---a path's {\em join style} augments the path's stroked region with a join region.  Round, bevel, and miter are the standard joins; PCL and XPS also support triangular joins.  Miter joins are subject to a {\em miter limit} so if a join is sufficiently sharp it exceeds the miter limit, the miter is either truncated or reverted to a bevel.  See Figure~\ref{fig:joins}.

\begin{figure}
\centering
\includegraphics[width=0.95\columnwidth]{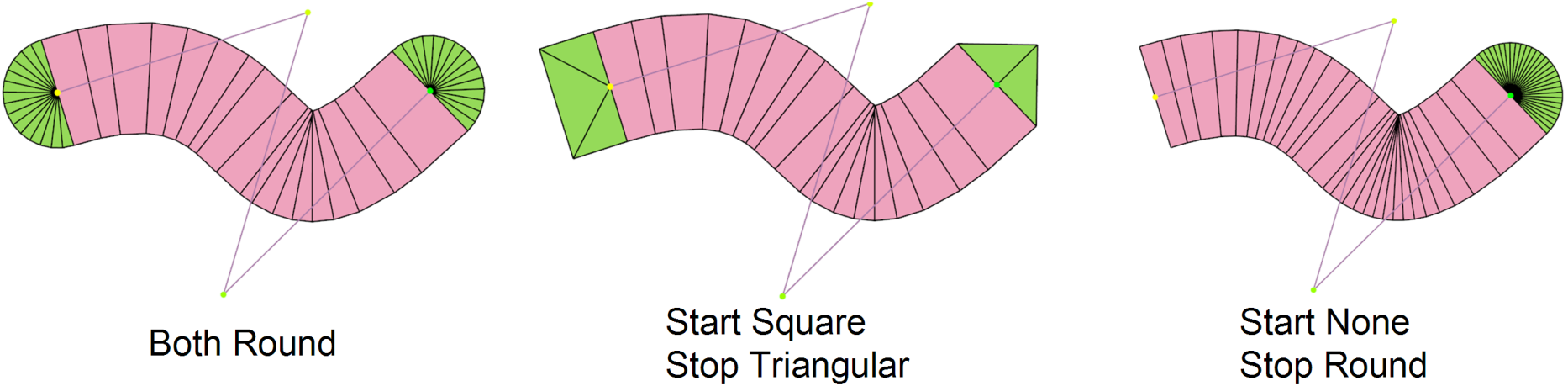}
\caption{\normalfont Tessellations of all supported cap styles. \label{fig:caps}}
\end{figure}

Caps are another way to augment the stroked region of a path. When a path segment does not join with another segment at its start or end, the stroked region beyond the unjoined start or end, respectively, can be augmented by a square or round cap.  PCL and XPS also support triangular caps.  See Figure~\ref{fig:caps}.


\subsubsection{Tessellation Approach for Caps and Joins}

We now discuss how to augment the tessellation process in Section~\ref{sec:TessProcess}
to support caps and joins with minimal modifications.  We treat caps and joins as zero-length segments---so the generator curve is a single point---yet with distinct start and stop tangent angles.  In other words, at a join or cap point $(x,y)$ there is a start tangent angle and stop tangent angle despite the fact that the generator curve's position is fixed at $(x,y)$ so has no actual gradient.  Think of this as a pivot.  Caps pivot 180\si{\degree} (similar to a cusp) while joins pivot based on the angle difference between the incoming and outgoing path segments meeting at the join.
These start and stop tangent angles depend on the path segments to which the join or cap connects.

For a join, the {\em start and stop} tangent angles are matched to the {\em stop and start} tangent angles of the incoming and outgoing path segments respectively.
The start and stop tangent angles at a start cap are matched to the capped path segment's start normal angle (tangent angle rotated +90\si{\degree}) and the angle 180\si{\degree} rotated from the normal angle by rotating counterclockwise (increasing angle) respectively.
The start and stop tangent angles at a stop cap are matched to the capped path segment's stop normal angle and the angle 180\si{\degree} rotated from the normal angle by rotating clockwise (decreasing angle) respectively.

For both joins and caps, the generator function is simply $\mathbf{g}(t) = (x,y)$ for the join/cap position; $M=1$; the sequence $p$ is trivially $p_0 = 0$ and $p_1 = 1$; and $\Delta_\Sigma(0) = 0$, $\Delta_\Sigma(1) = J$ where $J$ depends on the cap or join style, as will be discussed.

Specific to a join, the sequence $\Psi$ is $\Psi_0=\widehat{\nabla} \mathbf{g}_{a}(1)$ and $\Psi_1=\widehat{\nabla} \mathbf{g}_{b}(0)$ where $\mathbf{g}_{a}$ and $\mathbf{g}_{b}$ name the incoming and outgoing path segment generator functions; and $\delta_1 = \Psi_1 \ominus \Psi_0$.

Specific to a start cap, the sequence $\Psi$ is $\Psi_0=\widehat{\nabla} \mathbf{g}_{c}(0)$ and $\Psi_1=\Psi_0 + \pi$ where $\mathbf{g}_{c}$ names the cap path segment generator function; and $\delta_1 = +\pi$.

Specific to a stop cap, the sequence $\Psi$ is $\Psi_0=\widehat{\nabla} \mathbf{g}_{c}(1)$ and $\Psi_1=\Psi_0 + \pi$ where $\mathbf{g}_{c}$ names the cap path segment generator function; and $\delta_1 = -\pi$.

With this initialization established, we can specify how the various cap and join styles vary in their implementation.

For both caps and joins, respectively forcing either $r_N$ or $r_P$ to zero when $\delta_1$ is positive or negative ensures only the visible outside of the join is tessellated.  To ensure a watertight tessellation when forcing $r_N$ or $r_P$ to zero, we recommend introducing an extra quad that connects the adjacent path segment rib to the first or last rib of the join or cap.  These quads will typically be zero area, but avoid cracks from so-called T-junctions.

\subsubsection{Easy Joins: None, Miter, Triangular, Round}

The none, miter, and triangular join styles correspond to $J = 0, 1, 2$ respectively.

For a round join, compute $J = \ceil*{ \delta_1 / q }$ similar to Equation~\ref{eq:Delta_i} so that $q$ controls the tessellation quality for round caps just as it does for conventional curved path segments.  When the edges of the tessellated quads are small relative to a pixel, $q$ can be increased, thereby decreasing $N$, with no visible loss of round cap quality.

What makes these joins easy is they all have a constant distance $\frac{w}{2}$ from the join point to the outer rib vertex.

\subsubsection{Harder Joins: Miter Truncate and Revert}

With miter joins, there is not a constant distance from the join point to the miter vertices.  Set $J=3$ for the miter joins.  This generates three quads.  These three quads are sufficient to form the normal miter, the truncated miter, and a miter reverted to a bevel.  The details for how to compute miter vertices are standard stroking practice and beyond our scope.  Once computed with conventional methods, override rib vertices $\mathbf{P}_1$ and $\mathbf{P}_2$ when $\delta_1>0$ (or $\mathbf{N}_1$ and $\mathbf{N}_2$ when $\delta_1<0$) that otherwise are computed with Equations~\ref{eq:RibVertices-N} and \ref{eq:RibVertices-P}.

\subsubsection{Easy Caps: None, Triangular, Round}

The none and triangular cap styles correspond to $J$ being $0$ and $2$ respectively.

For round caps, same as round joins, set $J = \ceil*{ \delta_1 / q }$ again similar to Equation~\ref{eq:Delta_i} so that $q$ controls the tessellation quality for round caps just as it does for conventional curved path segments.  As with round joins when the edges of the tessellated quads are small relative to a pixel, $q$ can be increased with no visible loss of round cap quality.

\subsubsection{Harder Caps: Square Caps}

Square caps should set $J$ to 4 and then override $r_P$ when $\delta_1 > 0$ (or $r_N$ when $\delta_1 < 0$) to $\sqrt{2}$ for $j=1,3$ to push out to right angles these ``corner'' vertices to form a square cap.

\subsection{Complete Algorithm}
\label{sec:complete-algorithm}

Divide a path into links, one link per path segment, cap, and join.

\noindent For each link,

\noindent \hspace{1em} Compute $M$ and the sequences  $\Delta_\Sigma$, $p$, $\Psi$, $\delta$.

\noindent \hspace{1em} For $j=0...N$ where $N = \Delta_\Sigma(M)$:

\noindent \hspace{2em} Evaluate $\mathbf{g}(t(j))$ and $\mathbf{n}(j)$.

\noindent \hspace{2em} Generate rib vertices $\mathbf{N}_j$ and $\mathbf{P}_j$.

\noindent \hspace{2em} If $j>0$ emit the quad with vertices $\mathbf{N}_{j-1}$, $\mathbf{P}_{j-1}$, $\mathbf{N}_j$, $\mathbf{P}_j$.

\vspace{.5em}
\noindent
The algorithm's expressions evaluate equations in Sections~\ref{sec:theory} and \ref{sec:TessProcess}:
$\mathbf{g}(t)$ to one of the generator curve functions in Table~\ref{tab:path-segments}; $t(j)$ to Equation~\ref{eq:func_t}; $\mathbf{n}(j)$ to Equation~\ref{eq:func_N}.  $\mathbf{N}_j$ and $\mathbf{P}_j$ to Equations~\ref{eq:RibVertices-N} and \ref{eq:RibVertices-P}.

For the per-link intermediates, $M$ is the number of intervals (Section~\ref{sec:UnifiedTangentAngleIntervalRange}), $N$ is the last element index in the sequence $\Delta_\Sigma$ computed by Equation~\ref{eq:DeltaSigmaK};
sequences $p$, $\Psi$, and $\delta$ are computed by Equations~\ref{eq:p-value}, \ref{eq:Psi}, and \ref{eq:continuous-delta} respectively.

To visualize our complete algorithm's effectiveness, we present experiments in stroke rendering in an accompanying document \mycite{polar-stroking-robustness} where we apply our method to difficult and topologically varied path segments.  These experiments cover all the examples in our Figures~\ref{fig:loop} through \ref{fig:line_segment}, degenerate situations such as colocated and colinear control points, and the troublesome test cases found in our accompanying survey \mycite{path-stroking-survey}. 


\begin{figure}
\centering
\includegraphics[width=0.95\columnwidth]{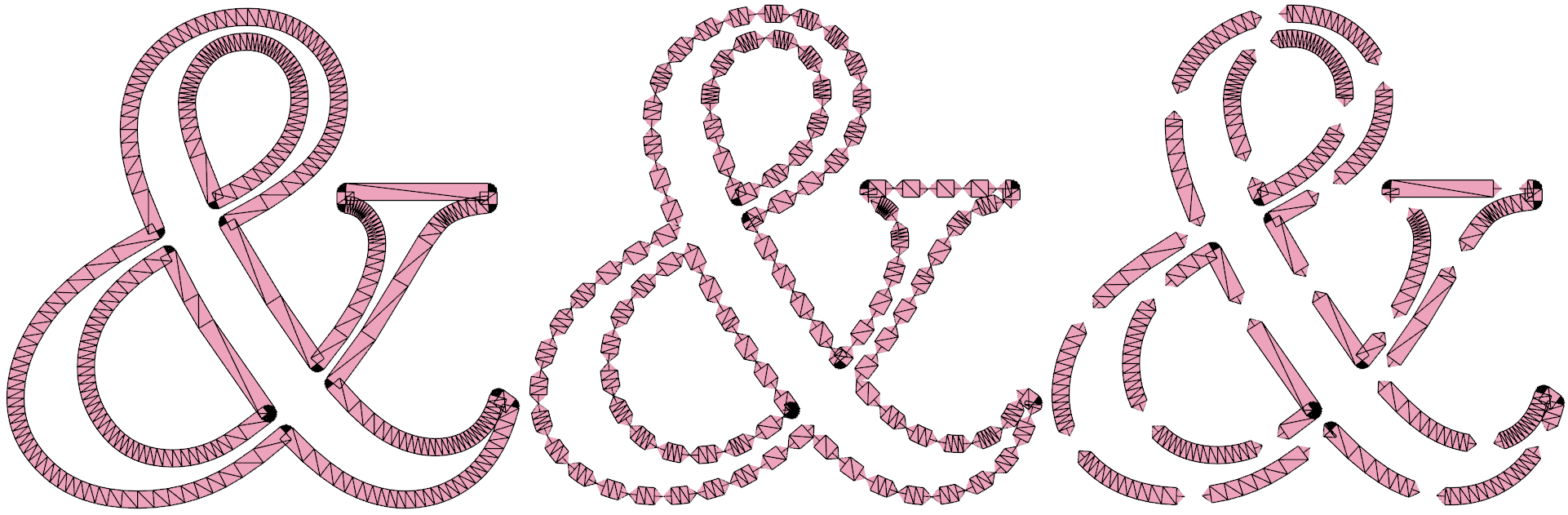}
\caption{\normalfont Stroked ampersand glyph with no dashing ({\em left}), then [1,1] ({\em center}) and [6,2]  ({\em right}) dashing  generated with polar stroking. \label{fig:dashing}}
\end{figure}

\section{Arc Length Along Paths}
\label{sec:arc_length}

Accurate algorithms for computing arc length along a curve typically rely on recursion
to build a chord length parameterization \cite{Gravesen1995,vincent2001fast}.  Our polar stroking method provides
a recursion-free way to build a chord length parameterization that adapts to curvature through its uniform steps in tangent angle.

With polar stroking, we estimate the arc length for a path segment as
\begin{align}
\int_{t=0}^{1} \, \left \lVert \mathbf{g'}(t)\right \rVert \, dt \approxeq \sum_{j=1}^{\Delta_\Sigma(N)} \left \lVert \mathbf{g}(t(j-1))-\mathbf{g}(t(j)) \right \rVert
\label{eq:arc-length-approximation}
\end{align}

The accuracy of this approximation depends on the tangent angle maximum step {\em q}.  The smaller the threshold angle for tangent angle step, the more accurate the approximation.
Floater~\shortcite{floater2005arc} provides a rationale for
why this kind of chordal parameterization of arc length for polynomials of degrees $\leq 3$, such as path rendering's $\mathbf{g}_C$ and $\mathbf{g}_Q$ forms, converges rapidly.

As our method provides a robust conversion of paths containing curved segments into strictly piecewise-linear sequences with bounded tangent angle changes, we expect our method to be useful in path-based NPR techniques
expecting piecewise-linear paths such as stroke parameterization \cite{schmidt2013stroke}.

\subsection{Cumulative Arc Length Texturing}
\label{cumulative-arc-length-shading}

Texture-based brush patterns \cite{Beach:1983:GST:800059.801141} are straightforward with a texture coordinate tracking cumulative arc length.

While tessellating a stroked path, we can accumulate the arc length and send this per-rib vertex pair as a texture coordinate for use by a fragment shader.  To apply a 2D texture, we would also send a second texture coordinate: 0 for the $\mathbf{N}_j$ vertex of the rib, and 1 for the $\mathbf{P}_j$ vertex.  Within a quad, we can reasonably linearly interpolate the arc length because polar stroking establishes the tangent angle change is bounded by $q$ within the quad.  
See Figure~\ref{fig:teaser}.C-F for examples.
For simplicity we do not mitigate texture folding artifacts, but techniques developed by Asente~\shortcite{Folding-Avoidance-In-Skeletal-Strokes} could apply.

\subsection{Dashing}

Dashing in path rendering breaks a path up into ``on'' and ``off'' sub-paths using the cumulative arc length along the path and a dash pattern and offset.
While splitting a line or circular arc is straightforward based on linear interpolation of parametric value or arc angle respectively, splitting curves
in the form of $\mathbf{g}_C$, $\mathbf{g}_Q$, and $\mathbf{g}_K$ is involved.

Using the polar stroking method to approximate such curves as a sequence of line segments with uniform steps in absolute tangent angle, we can quickly split
curves to determine cumulative arc length.  Figure~\ref{fig:dashing} shows our CPU-based dasher using polar stroking.

Rougier~\shortcite{rougier:hal-00907326} proposes a GPU shader-based approach to dashing, but to apply it path rendering assumes curved paths have been broken up into polylines.
Polar stroking would be a natural way to accomplish this.  The arc length texturing discussed in the prior subsection
pairs well with Rougier's method, enabling it to work on arbitrary paths.  We expect Rougier's method is just one of many applications
of arc length texturing on paths.
Yue et al.~\shortcite{yue2016function} is another example for cartography.


\begin{figure*}
\centering
\includegraphics[width=\textwidth]{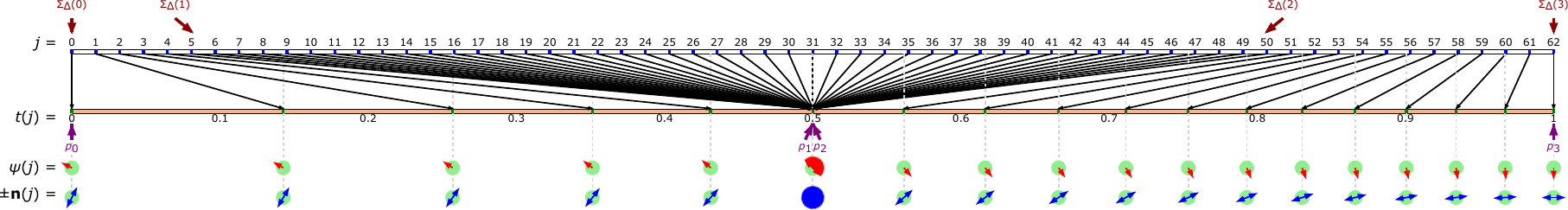}
\caption{\normalfont Tessellation chart for exact cusp segment from Figure~\ref{fig:cusp}.  The row of red arrows show the tangent angle at each $j$ stepping by 4\si{\degree} increments; the row of blue double arrows shows the normal vector at each $j$.  At the cusp $t=0.5$ observe the 180\si{\degree} of tangent angle change.  Each $j$ corresponds to a rib in the stroked tessellation of the cusp in Figure~\ref{fig:cusp}.}
\label{fig:cusp-chart}
\end{figure*}

\begin{table}
\caption{\normalfont Stroking grades for 21 stroking implementations from our supplemental survey, 8 rated A+  and consistent with polar stroking.}
\centering
\begin{tabular}{ | l || l | l |}
\hline
{\bf Stroking Implementation} &
{\bf Grade} &
{\bf Notes} \\
\hline
Acrobat Reader DC & {\bf \ \ A$+$} & \\
Cairo & {\bf \ \ B$-$} & \\
Chrome 74 (Windows) & {\bf \ \ B} & uses newer Skia \\
Direct2D CPU & {\bf \ \ A$+$} & \\
Direct2D GPU & {\bf \ \ A$+$} & \\
Firefox 66 (Windows) & {\bf \ \ A+} & uses Direct2D \\
Foxit Reader 8 & {\bf \ \ C} & \\
GSview 5.0 & {\bf \ \ D$-$} & \\
Illustrator CC 2019 & {\bf \ \ A$+$} & \\
Inkscape 0.91 & {\bf \ \ A$+$} & \\
Internet Explorer 11 & {\bf \ \ A$+$} & \\
MS Expression Design 4 & {\bf \ \ D} & \\
MS Office Picture Manager & {\bf \ \ A$-$} & \\
NV\_path\_rendering (NVpr) & {\bf \ \ A$+$} & OpenGL ext. \\
OpenVG 1.1 Reference Impl. & {\bf \ \ A$-$} & \\
Paint Shop Pro 7 & {\bf \ \ D$-$} & \\
PostScript (circa 1991) & {\bf \ \ C$+$} &  \\
Qt 4.5  & {\bf \ \ C$+$} &  \\
Skia CPU  & {\bf \ \ C} &  \\
Skia GPU  & {\bf \ \ C} & without NVpr  \\
SumatraPDF 3.1  & {\bf \ \ D} &  \\
\hline
\end{tabular}
\label{tab:survey-grades}
\end{table}

\section{Experimental Results}
\label{sec:experimental-results}

\subsection{Versus Uniform Parametric Tessellation }

Figure~\ref{fig:teaser}A compares
polar stroking ($q=4\si{\degree}$) of a cubic \Bezier cusp segment generating 67 quads with uniform parametric tessellation using Equation~\ref{eq:NaiveStroking} to also budget 67 quads.
While polar stroking generates
the expected double semicircle
tessellation of the cusp for any valid $q$, the uniform approach fails for any quad budget.

Figure~\ref{fig:teaser}B compares polar stroking ($q=4\si{\degree}$, $w=100$)  of a serpentine
cubic \Bezier segment with uniform parametric tessellation.  Each generates 126 quads.
Polar stroking generates a 69\% smaller maximum ordinary facet angle: 5.29\si{\degree} versus 17.12\si{\degree} for the uniform approach.
Polar stroking also has a similar mean and 76\% smaller standard deviation (3.79\si{\degree} versus
3.89\si{\degree} mean; 0.94\si{\degree} versus 3.99\si{\degree} SD)\@.

Crucially polar stroking provides a principled basis to determine how
many quads to tessellate, something uniform parameterization does not itself provide.

When seeding uniform tessellation to match the same quad output count as polar stroking,
we observe polar stroking to be superior at minimizing a path segment's ordinary facet angle maximum and standard deviation over wide variations of segment configuration, $q$, and stroke width.
This angle-based quality metric is scale-, translation-, and rotation-invariant and compares with a ground truth of 0\si{\degree} or ``infinite'' tessellation.
Our supplemental experiments \mycite{polar-stroking-robustness} provide further corroborating evidence
that polar stroking provides a statistically better facet angle distribution for the same number of quads.
This is likely due to polar stroking's similarity to chord length parameterization \cite{floater2005arc}
and our analysis \mycite{facet-angle} bounding ordinary facet angles to $2q$.  A thorough analysis is beyond our scope and left for future work.

\subsection{Polar Stroking of a Cusp}

To visualize why polar stroking forms the double semicircle cusp tessellation correctly for Figure~\ref{fig:cusp}'s stroked segment, Figure~\ref{fig:cusp-chart} charts polar stroking's tessellation method.
The chart shows how the integer values of $j$ driving the algorithm in Section~\ref{sec:complete-algorithm} generate $t(j)$ values to evaluate $\mathbf{g}_{C}(t)$ as well as generate the tangent angle $\psi(j)$ and normal $\pm\mathbf{n}(j)$.

The chart shows how polar stroking nonuniformly distributes the steps in $t$ (along the orange line).  Polar stroking forms the tessellated cusp  because values of $j \in [5..50]$
all map to the cusp at $t = 0.5$ but each $j$ has a distinct normal advanced in uniform tangent angle steps.

This chart is just one from 27 carefully-curated stroking examples in our supplemental
materials \mycite{polar-stroking-robustness} to demonstrate experimentally
why polar stroking operates robustly.

\subsection{Comparison to Stroking Implementations}

\begin{figure}
\centering
\includegraphics[width=0.86\columnwidth]{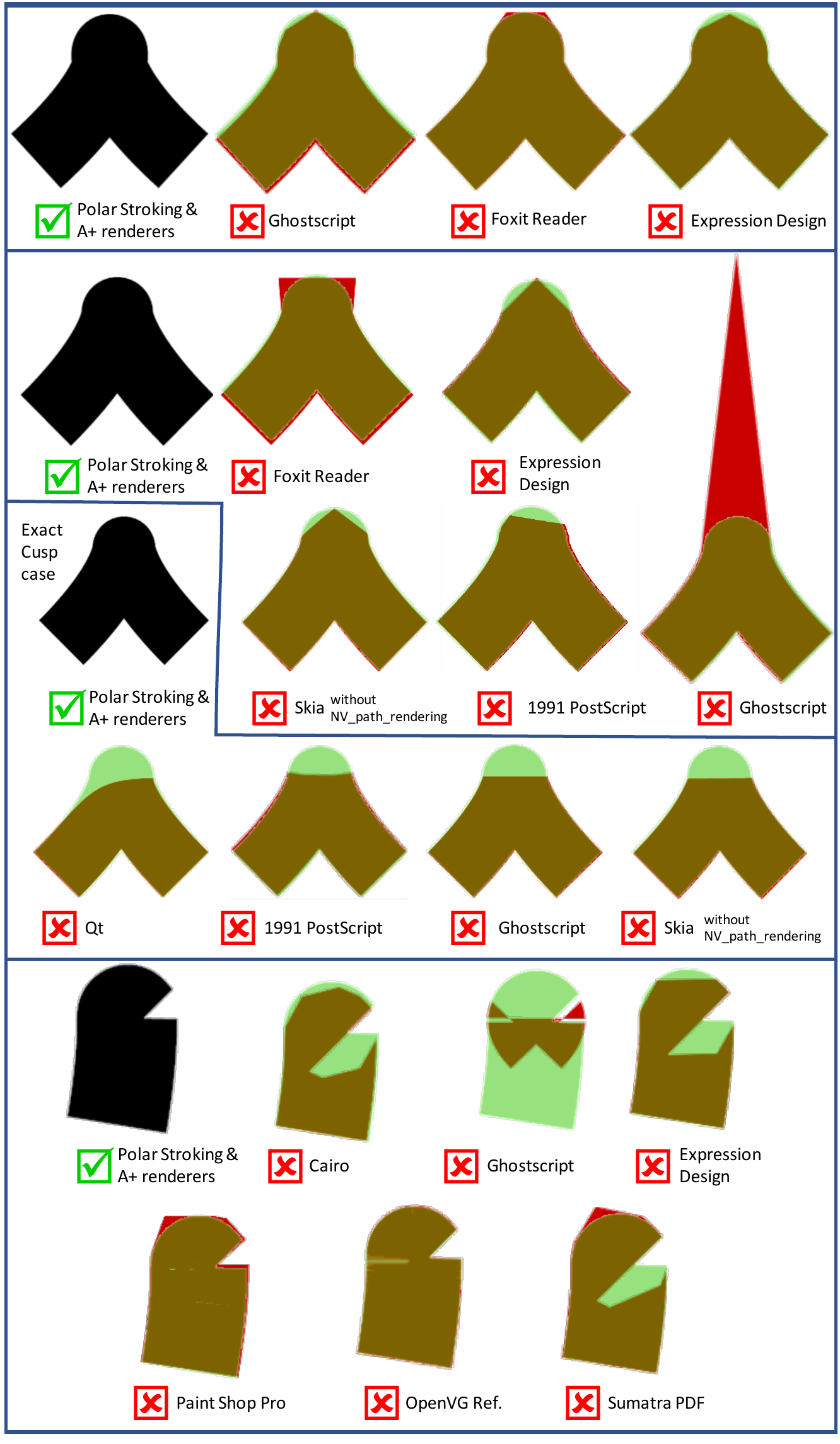}
\caption{\normalfont Survey of real-world stroking results showing divergences from polar stroking. Olive: mutual overlap of polar stroking and the other stroking implementation.  Red: other implementation's excess coverage.  Green: absent coverage.
Note: image alignment is inexact.
}
\label{fig:survey-images}
\end{figure}

Table~\ref{tab:survey-grades} summarizes our supplemental survey \mycite{path-stroking-survey} listing grades we assigned to real-world vector graphics implementations for stroking quality.
Figure~\ref{fig:survey-images} collects results from the survey noticeably different from what polar stroking theory expects.
We emphasize A+ means rendering matched {\em both} polar stroking theory and the survey's best consensus.
See our supplemental survey for complete discussion and images.


\section{Limitations}
\label{sec:limitations}

We collect and amplify limitations of our methods and offer advice:
\begin{itemize}
\item As $q$ diminishes, more tessellated quads will be generated; vice versa, if $q$ is insufficiently small, facet angles will be noticeable.  Assessing how varying $q$ affects pixel quality is left for future work.
\item To guarantee a facet angle bound of $\theta$, $q$ should be $\theta / 2$ (Section~\ref{sec:TessProcess}).
Ribs immediately bracketing a rib at an inflection point on cubic \Bezier segments have larger facet angles (closer to $\theta$) to offset the zero or near-zero facet angle at the inflection point.
 Our future work provides a tighter bound.
\item Our tessellation method works properly when bow-tie quads are rasterized as such (Section~4.1).  Treating bow-tie quads as overlapping triangles, as GPUs do by default, exaggerates stroked coverage for bow-tie quads; the excess coverage diminishes as $q$ diminishes.  When paths are stroked with round joins and use either round caps or always closed contours, the caps and joins hide the excess coverage.
\item When steps by $q$ are minuscule, solving for $t$ in Equation~\ref{eq:angle2t} may not guarantee strictly increasing $t$ values as $j$ increases.
``Stencil-then-cover'' methods update pixels once per path so hide this negligible misordering.
For arc length computations, Equation~\ref{eq:arc-length-approximation} can accommodate this by zeroing step distances when $t(j-1) > t(j)$.
\item Our arc length approximation (Equation~\ref{eq:arc-length-approximation}) converges fast when $q$ diminishes but is biased to underestimate the ideal arc length.  As both arc length texturing and dashing rely on such arc length computations, these methods are more accurate when $q$ is diminished.
\item Our method does not consider the scale of pixels. Heuristics could boost $q$ (reducing the tessellation) for segments, caps, and joins near or below the scale of pixels to avoid excessive tessellation relative to the available pixels to cover.\
\item Our facet angle and tangent angle step bounds---dependent on $q$---are not be maintained after non-conformal transformations (i.e., nonuniform scaling, shearing, or projection) of the tessellation.  Our future work addresses this.
\end{itemize}


\section{Conclusions}
\label{sec:conclusion}

Prior to our work, vector graphics lacked a principled theory of path stroking consistent with the expectations
of established path rendering standards.  Leveraging our new theory of stroking---based on offset curves parameterized by tangent angle---we developed
a novel method to
tessellate stroked paths by taking uniform steps in tangent angle magnitude.  Our method intuitively bounds
the tessellation error at facet angles and matches the number of quadrilateral primitives generated to each path segment's absolute tangent rotation while robustly handling cusps, inflections, general conics, and degenerate segments---all without recursion, so GPU-amenable.

We made sure to harmonize our theory and methods with practical requirements of modern path rendering standards.
We explained how our theory and method
extends to handle caps and joins.  Our theory and methods make approximating the cumulative arc length of
paths straightforward, even for lengthy paths with high curvature, and we leveraged this ability to implement dashing and arc length texturing methods.

Accompanying this paper are sample images for quality evaluation and
source code for a path stroking testbed that demonstrates our method
using the CPU to perform the stroked tessellation.  While beyond the
scope of this paper, we have also successfully implemented our methods
on modern programmable GPUs.


\begin{acks}
Sanjana Wadhwa assisted validating the polar stroking method and its initial GPU implementation.
\end{acks}

\pdfbookmark[1]{References}{bkmk:references}

\bibliographystyle{ACM-Reference-Format}


\appendix


\section{Path Filling Theory in Brief}
\label{sec:fill-theory}

By representing the pixel location $(x,y)$ as a number $w = x + i y$ on the complex plane and the path as a contour defined by a closed complex function $\gamma$, complex analysis provides a means to compute an integer {\em winding number} of $w$ with respect to $\gamma$ expressed as a contour integral
\[
n(\gamma,w) = \frac{1}{2\pi i} \oint_{\gamma}  \frac{dz}{z - w}
\]
where $|n(\gamma,w)|$ measures the whole number of times that the contour $\gamma$ ``winds around'' $w$ while the sign of $n(\gamma,w)$ indicates whether the winding is counterclockwise when $n(\gamma,w)>0$, clockwise when $n(\gamma,w)<0$, or not wound within when $n(\gamma,w)=0$.  A path $P$ in vector graphics can specify $m$ contours, say $\gamma_1$ through $\gamma_m$, so the net winding number of $w$ with respect to $P$ is
\[
n(P,w) = \frac{1}{2\pi i} \sum_{i=1}^{m} \oint_{\gamma_i}  \frac{dz}{z - w}
\]
Whether a pixel at $(x,y)$ is within $P$ is decided by one of two standard support predicates
\begin{align*}
P_{\text{nz}}(x,y) &=
\begin{cases}
1, \text{ if } n(P,x + i y)\neq 0 \\
0, \text{ otherwise}
\end{cases}
\\
P_{\text{eo}}(x,y) &=
\begin{cases}
1, \text{ if } n(P,x + i y) \, \mathbf{ mod } \,  2\neq 0 \\
0, \text{ otherwise}
\end{cases}
\end{align*}
$P_{\text{nz}}$ and $P_{\text{eo}}$ respectively correspond to the standard {\em nonzero} and {\em even-odd} fill rules in path rendering systems
so the winding number concept is explicit in how path filling is specified.

\end{document}